\documentclass{article}
\usepackage{paper}
\usepackage{float}

\usepackage{booktabs}
\usepackage{tabularx}
\usepackage{makecell}
\usepackage{siunitx} 
\usepackage{placeins}

\begin{document}

\title{Design rules for carboborothermic reduction synthesis of high uranium density \ce{UB4}–\ce{UBC} composites}

\author[1, 2]{Riley Moeykens}
\author[1, 2]{Anthony Albert-Harrup}
\author[1, 2]{David Simonne}
\author[3]{Mehmet Topsakal }
\author[1, 4]{Ericmoore Jossou\thanks{Corresponding author, ejossou@mit.edu}}

\affil[1]{Materials in Extreme Environments Laboratory, Massachusetts Institute of Technology, 77 Massachusetts Ave, Cambridge, 02139, MA, USA}
\affil[2]{Department of Nuclear Science and Engineering, Massachusetts Institute of Technology, 77 Massachusetts Ave, Cambridge, 02139, Massachusetts, United States}
\affil[3]{Department of Nuclear Science and Technology, Brookhaven National Laboratory, Upton, 11973, New York, United States}
\affil[4]{Electrical Engineering and Computer Science, Massachusetts Institute of Technology, 77 Massachusetts Ave, Cambridge, 02139, Massachusetts, United States}

\maketitle

\begin{abstract}
Uranium borides are promising candidate fuel forms for use in advanced nuclear reactors due to their high thermal conductivity and potential for dual use as both fuel and burnable absorber.  In this work, uranium tetraboride (\ce{UB4}) and uranium monoboroncarbide (\ce{UBC}) composite were synthesized by using industrially scalable carboborothermic  reduction method. \color{blue}The final uranium boride phase composition is sensitive to the sample holding crucibles (\ce{Al2O3} and graphite) such that graphite supply excess carbon, promoting the formation of a predominant UBC phase. The high temperature \textit{in situ} synchrotron X-ray diffraction of pristine \ce{UB4}–\ce{UBC} show persistence \ce{UB4}, UBC and \ce{UO2} phases while preoxidized \ce{UB4}–\ce{UBC} leads to predominant \ce{UB2O6} and \ce{U3O8} formation due to progressive oxidation and boron loss at high temperature. The oxidation behavior was further characterized using thermogravimetric analysis, \color{black}allowing direct comparison with other potential accident tolerant fuels such as \ce{UB2}, \ce{U3Si2}, UC and UN. The \ce{UB4}–\ce{UBC} shows higher uranium loading than monolithic \ce{UB4} and demonstrates promising oxidation behavior at high temperature, pointing to its potential as an improved uranium boride-based fuel form. 
\end{abstract}


\section{Introduction}

In 2011, the severe accident at Fukushima Daiichi Power Station pointed to the limitations of the conventional uranium dioxide-zirconium fuel-cladding system under severe accident scenarios. 
This event motivated renewed research into advanced fuel concepts leading to international efforts in the development of accident tolerant fuels (ATFs). 
The U.S. Congress and the Department of Energy’s Office of Nuclear Energy launched the ATF program with the objective of developing non-traditional fuel and cladding systems that can withstand accident scenarios, such as loss-of-coolant events lasting longer than \qty{1}{\hour} \autocite{Carmack2013-bs}.
Such fuels must show improved performance during normal operations, transients, design-basis and beyond design-basis accidents, as well as compatibility with non-traditional coolant and varied neutron-spectra.

Many of the proposed long-term ATF concepts, such as \ce{UB2}, UC, UN and \ce{U3Si2} fuels, have performance, and safety advantages over \ce{UO2} such as improved thermal conductivity and higher fissile density \autocite{IAEA2008Thermophysical, GONZALES2021153026, KARDOULAKI2020153216, JOSSOU201741}. However, there is limited understanding of their thermophysical properties, pellet-cladding chemical and mechanical interactions, and irradiation behavior. With recent improvements in advanced computing \autocite{Faibish2021AFQ, TERRANI2020152267}, throughput irradiation facilities \autocite{Bhowmik2023MultiLevelIrradiation}, and advanced characterization tools, such as synchrotron-based X-ray scattering and spectroscopy techniques \autocite{DEGUELDRE2016242, Lang2015, THOMAS2020152161}, the acceleration in qualification of ATFs is now more feasible than ever. 

In particular, uranium borides are of interest as ATFs for use in light water reactors (LWRs) and high temperature gas reactors (HGTRs). 
These materials are high-temperature refractory metalloids with a high thermal conductivity, improved thermal stability \autocite{Koenig1960CeramicCermet}, and with potential for use as a burnable absorber \autocite{KARDOULAKI2020153216} or as a composite with \ce{UO2} or \ce{U3Si2} \autocite{JOSSOU201741, TURNER2020151919, WATKINS2022153502, Faibish2021AFQ}. 

The high thermal conductivity reduces the center-line fuel temperature, enhancing tolerance to operational transients and loss-of-coolant accident conditions. Although \ce{UB2} and \ce{UB4} appear to have improved thermophysical properties for use as nuclear fuels, the high temperature structural stability, oxidation behavior and irradiation behavior at temperatures and conditions relevant to accident scenarios are not well known. 
Previous work has reported an oxidation onset temperature as high as \qty{500}{\degreeCelsius} for \ce{UB4} \autocite{Guo2019}. 
While \ce{UB2} has improved uranium loading compared to \ce{UO2}, the sintering temperature is demonstrated to be as high as \qty{1800}{\degreeCelsius} based on recent studies \autocite{TURNER2020152388}. Meanwhile, \ce{UB4} and uranium monoboroncarbide (\ce{UBC}) can be sintered at lower temperatures \cite{Guo2019}, making them attractive alternatives from a manufacturing perspective, particularly if comparable thermophysical performance can be achieved with reduced processing energy.

Although \ce{UB4} demonstrates favorable thermomechanical properties, its uranium density of \qty{7.94}{\g\per\cm^3} \autocite{MENOVSKY1984519} is lower than that of \ce{UO2} (\qty{9.67}{\g\per\cm^3}) \autocite{BURR201945}. 
In comparison, \ce{UBC} has a higher uranium loading (\qty{\approx10.9}{\g\per\cm^3}), suggesting that \ce{UB4}–\ce{UBC} composites could achieve uranium densities similar to \ce{UO2} while maintaining the beneficial characteristics of uranium borides.

Furthermore, only a few documented studies investigate the air oxidation behavior of \ce{UB4} and \ce{UBC}. It has generally been observed that while actinoid metal borocarbides have high oxidation resistance in dry air, they rapidly hydrolyze in moisture \autocite{Rogl1990ActinoidmetalBoronCarbides}. 
Therefore, we expect that the formation of a \ce{UB4}–\ce{UBC} composite will improve the oxidation resistance while also increasing uranium loading. 

The large scale fabrication of these borides composite have to be done using a scalable industrial fabrication technique. Uranium borides have been fabricated by arc melting \autocite{KARDOULAKI2020153216, TURNER2020152388, KARDOULAKI2021152690,Guo2019}, suitable for research purpose due to the small batch processing. Meanwhile, the carboborothermic  reduction route is industrially scalable but requires optimization of several parameters, including stoichiometry of the starting feedstock, sintering temperature, duration, and atmosphere. The fabrication of \ce{UB2} via carboborothermic  reduction has been demonstrated at \qty{1800}{\degreeCelsius} for \qty{1}{\hour} \autocite{TURNER2020152388}, while \ce{UB4} has been synthesized using the same approach at \qty{1500}{\degreeCelsius} under flowing argon \autocite{Guo2019}, however at relatively lower uranium loading. 

The sintering of a \ce{UB4}–\ce{UBC} is expected to occur at relatively lower sintering temperature compared to \ce{UB2} but with comparable uranium loading and improved oxidation behavior. \color{blue}Therefore, the main purpose of this work is to systematically optimize the fabrication of \ce{UB4} and \ce{UB4}–\ce{UBC} composites using an industrially viable  solid state sintering method at temperatures lower than \ce{UB2} fuel. Thermodynamic calculations is used to determine the sintering temperature range for specific starting feedstock stoichiometry. \color{black}The high temperature structural stability and oxidation behavior of the fuel is investigated using scanning electron microscope (SEM), energy dispersive X-ray (EDX) imaging, \textit{in situ} high temperature synchrotron X-ray diffraction (SXRD) measurements, and thermal gravimetric analysis (TGA). 

\section{Methodology}

\subsection{\textbf{\ce{UB4} synthesis }}

The initial feedstocks consisted of  \color{blue}\ce{UO2} (\qty{99.99}{wt.\percent} purity,  \color{black}\qtyrange{1}{3}{\um} particle size) from Ibilabs Inc, \ce{B4C} (\qty{99.99}{wt.\percent} purity, \qty{45}{\um}), and graphite (\qty{99.99}{wt.\percent}, \qty{297}{\um}) from Naografi. Powder handling was performed in a double-sided UNIlab Pro MBraun glovebox under high-purity argon (\qty{99.99}{\percent}), with oxygen and moisture levels maintained below \qty{0.5}{\ppm}. The synthesis of \ce{UB4} proceeded via the following reaction \autocite{Guo2019}: 

\begin{equation}
    \mathrm{UO_2(s) + B_4C(s) + C(s) \rightarrow UB_4(s) + 2CO(g)}
    \label{eq:UB4Synthesis}
\end{equation}

Six samples were prepared using the same precursor stoichiometry and were ball-milled for \qty{4.5}{\hour} in a tungsten jar with stainless steel balls at rotational speeds of \qtyrange{300}{400}{\rpm}, ensuring comparable particle size distributions. Each mixture was pressed into green pellet under 300–400 MPa. The resulting green pellets had masses of approximately \qtyrange{0.5}{2}{\g} and  \color{blue}thicknesses of about 2 mm. \color{black}The samples were sintered in an alumina-tube furnace under flowing argon at \qty{100}{\SCCM} (Standard Cubic Centimeters per Minute) at a \qty{5}{\degreeCelsius \per \minute} heating rate with two \qty{30}{\minute} isotherms at \qty{500}{\degreeCelsius} and \qty{1000}{\degreeCelsius}. 

The experimental parameter space for sintering single phase \ce{UB4} was systematically explored by varying sintering environment, temperature, and time. All \ce{UB4} samples were prepared using the identical pre-sinter stoichiometry of \ce{UO2}:\ce{B4C}:C which is equivalent to 1.00:1.04:1.00, with an excess of \ce{B4C} to account for volatilization during heating. To further show the effects of the sintering crucible, a stochiometric mixture of \ce{UO2}, \ce{B4C}, and C was ball-milled for \qty{2.5}{\hour} at 330 rpm. Two samples were pressed into pellets and sintered simultaneously at 1500\textsuperscript{o}C for \qty{1}{\hour} under \qty{100}{\SCCM} flowing argon—one in a graphite crucible and the other in an alumina crucible. 


\subsection{\textbf{\ce{UB4}–\ce{UBC} synthesis }}
The synthesis of \ce{UB4}–\ce{UBC} is based on the partial reduction of the starting feedstock with unreacted \ce{UO2} and excess carbon from the graphite crucible as shown in Equation (8):
\begin{equation}
    10\mathrm{UO}_2(s) + 5\mathrm{B}_4\mathrm{C}(s) + 19\mathrm{C}(s) \rightarrow 4\mathrm{UB}_4(s) + 4\mathrm{\ce{UBC}}(s) + 20\mathrm{CO}(g) + 2\mathrm{UO}_2(s)
\end{equation}
During preliminary efforts to synthesize \ce{UB2}, \ce{UB4}–\ce{UBC} composite samples were also formed. Subsequent analysis of these trials enabled the determination of effective stoichiometry and sintering parameters for the deliberate synthesis of \ce{UB4}–\ce{UBC}. The \ce{UB4}–\ce{UBC} composite sample was produced from a feedstock mixture of graphite, uranium dioxide and boron carbide that had been ball milled in a tungsten milling jar at a speed of 300 to 400 rpm for 2 to 4.5 h. The samples were sintered at \qty{1450}{\degreeCelsius} to \qty{1700}{\degreeCelsius} under\qty{100}{\SCCM} argon flow in a graphite crucible. \color{blue}The samples were sintered under flowing argon gas, although the oxygen partial pressure was not quantified during the experiment. The addition of a Zr oxygen getter further suppressed the formation of deleterious boride oxides, as confirmed by XRD results. \color{black}

\subsection{Oxidation experiments}

Oxidation experiments were conducted by following two different routes at temperatures ranging from room temperature to \qty{900}{\degreeCelsius}: (I) pre-air oxidation in a tube furnace for \textit{ex situ }and\textit{ in situ} SXRD analysis and (II) thermogravimetric analysis of weight change due to oxidation in synthetic air (21.49 \% O\textsubscript{2 }in N\textsubscript{2}). 
For SXRD characterization, small quantities of the sample, either in the form of a coarse powder or as a small piece of a sintered pellet with a mass on the order of tens of milligrams, were placed into small alumina crucibles which were then placed inside a larger alumina crucible positioned at the center of the furnace. 
The \ce{UB4} samples were oxidized in flowing compressed air at flow rate of \qty{\approx 100} sccm at \qty{350}{\degreeCelsius} and \qty{900}{\degreeCelsius} while the \ce{UB4}–\ce{UBC} samples were oxidized in moist air under laboratory ambient conditions, where exposure to atmospheric humidity was unavoidable, at temperatures of \qty{300}{\degreeCelsius} and \qty{900}{\degreeCelsius}. \color{blue}Due to the time required for the in situ SXRD experiments, detailed analysis was performed on one representative pellet for each composition. As such, the results should be interpreted as representative rather than statistically exhaustive.\color{black}

Thermogravimetric analysis (TGA) was performed using a TA Instruments Discovery TGA 5500 at the MIT Institute for Soldier Nanotechnologies (ISN). The instrument operates over a temperature range of \qtyrange{40}{1000}{\degreeCelsius} with a temperature precision of \qty{\pm0.005}{\degreeCelsius} and a mass resolution of \qty{0.1}{\micro\g}.
Oxidation behavior was evaluated under flowing synthetic air at \qty{25}{\mL\per\minute} using a temperature program consisting of an equilibration at \qty{50}{\degreeCelsius} followed by a linear ramp of \qty{5}{\degreeCelsius\per\min} to \qty{900}{\degreeCelsius}. 
High-purity alumina crucibles (\qty{6.8}{\mm} outer diameter, \qty{4}{\mm} height, \qty{\approx 300}{\mg}; MSE Supplies) were first subjected to the same thermal and gas flow conditions to establish a baseline and account for background mass changes. 
Subsequently, \qtyrange{2}{3}{\mg} of uranium boride powder was loaded into the crucibles for thermal analysis. 
Baseline correction was performed by applying a spline fit to the empty-crucible data and subtracting this background from the sample measurements over the entire temperature range, enabling accurate determination of oxidation-induced mass changes.

\subsection{SEM imaging}

SEM and EDX images were collected to determine the microstructure, the elemental composition, and homogeneity after sintering. The Zeiss Sigma \num{300} VP Field Emission SEM located at the MIT ISN was used for imaging. 
A voltage of \qty{25}{\kV} was used for the EDX data collection, to ensure that the electron energy was well above the uranium L-\(\alpha\) edge at \qty{13.6}{\keV}. 

\subsection{Synchrotron X-ray diffraction measurement and analysis}

\begin{figure}[H]
    \centering
    \includegraphics[width=1.0\linewidth]{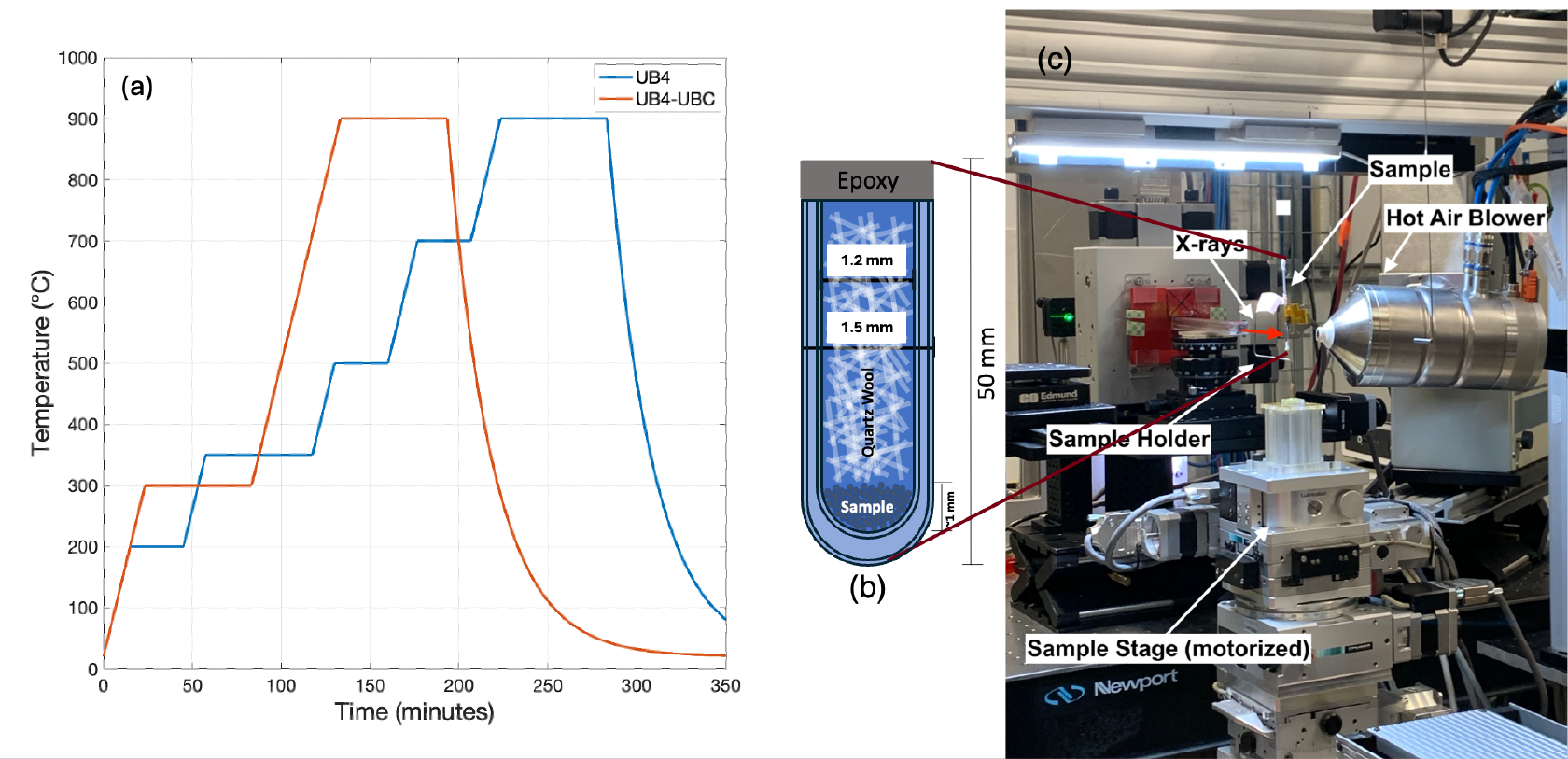}
    \caption{
    (a) Heating profile for synchrotron characterization of a \ce{UB4} and \ce{UB4}–\ce{UBC} composite sample 
    (b) Double layer containment sketch for \ce{UB4}–\ce{UBC} sample in quartz capillaries 
    (c) High temperature in situ heating XRD setup at 28-ID at NSLS II.
    }
    \label{fig:Figure1}
\end{figure}

XRD measurement experiments were conducted at the 28-ID-2 beamline at the National Synchrotron Light Source-II (NSLS II) at Brookhaven National Laboratory (BNL). The oxidation conditions, sample preparation, and heating profile during measurement at the 28-ID-2 beamline varied between the \ce{UB4} and \ce{UB4}–\ce{UBC} samples. 
The differing heating profiles are illustrated in Figure \ref{fig:Figure1}a.

In preparation for the XRD measurements at the 28-ID-2 beamline at NSLS II, powder samples were loaded into quartz capillaries to meet the beamline radiological containment requirements. Due to the radioactive nature of the sample, all samples were doubly contained using two concentric quartz capillaries to mitigate the risk of contamination in the event of breakage during the experiments as shown in Figure \ref{fig:Figure1}b. 

The dimensions of the quartz capillaries and sealants varied slightly between experiments. For the \ce{UB4}–\ce{UBC} composites, samples were sealed in double-walled quartz capillaries with closed, rounded bottoms. The inner capillary had an outer diameter of 1.2 mm and a wall thickness of 0.1 mm, while the outer capillary had an outer diameter of 1.5 mm and a thickness of 0.1 mm. Both capillaries were approximately 50 mm in height. Samples were initially ground with a mortar and pestle to reduce particle size and to ensure homogeneous mixing of the feedstocks. A thin powder layer of approximately 1 mm was loaded into the inner capillary, followed by the insertion of quartz wool to compact the sample and restrict its movement. The inner capillary containing the sample was then inserted into the outer capillary and sealed with high-temperature epoxy resin which is stable up to \qty{1300}{\degreeCelsius}. The epoxy was allowed to cure fully before packaging and shipment to the beamline.

For the \ce{UB4} measurement, a different double-capillary configuration was used. These capillaries have outer diameters of 2.1 mm and 1.9 mm, respectively, with thickness of 0.2 mm with open end that allows the mounting of an alumina rod with outer diameter of  \qty{\approx1.8} mm inside the capillary to provide a physical buffer between the sample and the sealant. The alumina rod, inner capillary, and outer capillary were sealed together using alumina-based glue. After drying, an approximately 1 mm-thick layer of sample was loaded directly above the alumina rod and compacted with quartz wool and a tertiary quartz capillary. A final layer of alumina glue was applied to seal the assembly from the top.

The experimental setup is schematically shown in Figure \ref{fig:Figure1}c.
Samples were heated to the desired temperatures using a hot air blower.
A two-dimensional (2D) CsI flat panel area detector (Dexela 2923) with a pixel size of \qtyproduct{75x75}{\um} was mounted orthogonally to the X-ray beam to collect the diffraction patterns. 
The wavelength of the incident X-ray beam is \qty{0.1811}{\angstrom} (\qty{68.46}{\keV}). 
The sample-to-detector distance was calibrated using a NIST LaB\textsubscript{6} standard and determined to be \qty{0.8}{\m}.
The beam spot size was \qtyproduct{250x250}{\um}. 
The 2D diffraction patterns were cleared from electronic artifacts created by dead pixels and from the shadow of the beam stop. 
The detector acquisition time was set at \qty{0.1}{\second}.
To improve the counting statistics while avoiding saturation, multiple frames were acquired and averaged. 
Meanwhile, the laboratory XRD (LXRD) measurements were performed using a Siemens D5000 diffractometer with Cu K-$\alpha$) radiation source ($\lambda = \qty{1.5418}{\angstrom}$), with $2\theta$ ranging from ranging from \ang{20} to \ang{60} a $2\theta$ step size of \ang{0.02}, and an acquisition time of \qty{1}{\second} per step.

The data analysis is automated at 28-ID-2, such that the 2D diffraction images are masked and integrated to produce the corresponding 1D diffraction patterns. 
The XRD patterns were indexed using JADE Pro (Materials Data, Inc.) \autocite{JADEPro} with access to the International Centre for Diffraction Data (ICDD) Powder Diffraction File (PDF)-5+ database \autocite{Kabekkodu2024PDF5Plus}. 
All the XRD data analysis was performed using the open-source General Structure Analysis System II (GSAS-II) software package \autocite{Toby:aj5212}. 
A module in GSAS-II referred to as \texttt{GSASIIscriptable} allows for automation of refinement and is utilized in this work. 
The (110) and (200) peaks of alumina, resulting from the alumina rod at the base of the sample, were excluded from the fitting. 

\section{Results}

\subsection{Gibb’s free energy calculations}

The Gibbs free energy of reaction was calculated to show the correlation between the partial pressure of CO gas released, temperature, and spontaneity of the reactions. The results serve as guideline for choosing the sintering temperatures.  The Gibbs free energy for a specific reaction can be computed based on the molarity of reactants and products using Equation (2):

\begin{equation}
    \Delta G_{\mathrm{rxn}}(T) = \sum_{p} c_{p}\,\Delta G_{f,p}(T) - \sum_{r} c_{r}\,\Delta G_{f,r}(T)
\end{equation}


\(\Delta\)G\textsubscript{rxn }is the Gibbs free energy of formation for a reactant (r) or product (p) in units of J/mol at standard pressure, C is the molar quantity of each reactant or product and T is the temperature. When gaseous species are present in the reactants or products, the Gibbs free energy of the reaction can be written with respect to the partial pressures of gases produced in the reaction. In the case of the synthesis of \ce{UB4}, CO is produced as a by-product of the carboborothermic  reduction reaction:

\begin{equation}
    \mathrm{UO_2(s) + B_4C(s) + C(s) \rightarrow UB_4(s) + 2CO(g)}
\end{equation}

The effect of the CO partial pressure on the Gibbs free energy of reaction can be written as\autocite{TURNER2020152388}: 

\begin{equation}
    \Delta G_{\text{rxn, pressure}} = C_{\mathrm{CO}} \, RT \ln\left(\frac{P_{\mathrm{CO}}}{101 \, \text{kPa}}\right)
\end{equation}

R is the ideal gas constant given as \qty{8.314}{\joule \per \kelvin \per \mol}, C\textsubscript{CO}=2 for the synthesis of \ce{UB4} and P\textsubscript{CO }is the partial pressure of CO in kPa. Therefore, for the synthesis of \ce{UB4}, the Gibbs free energy of reaction may be given as:

\begin{equation}
    \Delta G_{\mathrm{rxn}}\left(T, P_{\mathrm{CO}}\right)
    = C_{\mathrm{CO}}\,RT \ln\left(\frac{P_{\mathrm{CO}}}{101\,\mathrm{kPa}}\right)
    + \Delta G^\circ(T)
\end{equation}
\(\Delta{G}^o(T)\) is the standard Gibbs free energy change of the reaction at temperature T. The temperature-dependent Gibbs free energy of formation was extracted from the thermochemical data of pure substances for \ce{UO2}, \ce{B4C}, C, CO, and \ce{UB4} \autocite{Barin1989ThermochemicalData}. This data was fitted to a first-degree polynomial, then extrapolated to the temperature range used during synthesis. It is hypothesized that the formation of a \ce{UB4}–\ce{UBC} composite from a \ce{UO2}, \ce{B4C}, and C feedstock proceeds via the following overall reaction in a graphite crucible: 

\begin{equation}
    5\mathrm{UO}_{2}(s) + 2\mathrm{B}_{4}\mathrm{C}(s) + 12\mathrm{C}(s) \rightarrow \mathrm{UB}_{4}(s) + 4\mathrm{\ce{UBC}}(s) + 10\mathrm{CO}(g)
\end{equation}

This is potentially a two-step reaction where the first is given by Equation (6) and the second is given as:

\begin{equation}
    4\mathrm{UO}_2(s) + \mathrm{B}_4\mathrm{C}(s) + 11\mathrm{C}(s) \rightarrow 4\mathrm{\ce{UBC}}(s) + 8\mathrm{CO}(g)
\end{equation}

\begin{figure}
    \centering
    \includegraphics[width=1.0\linewidth]{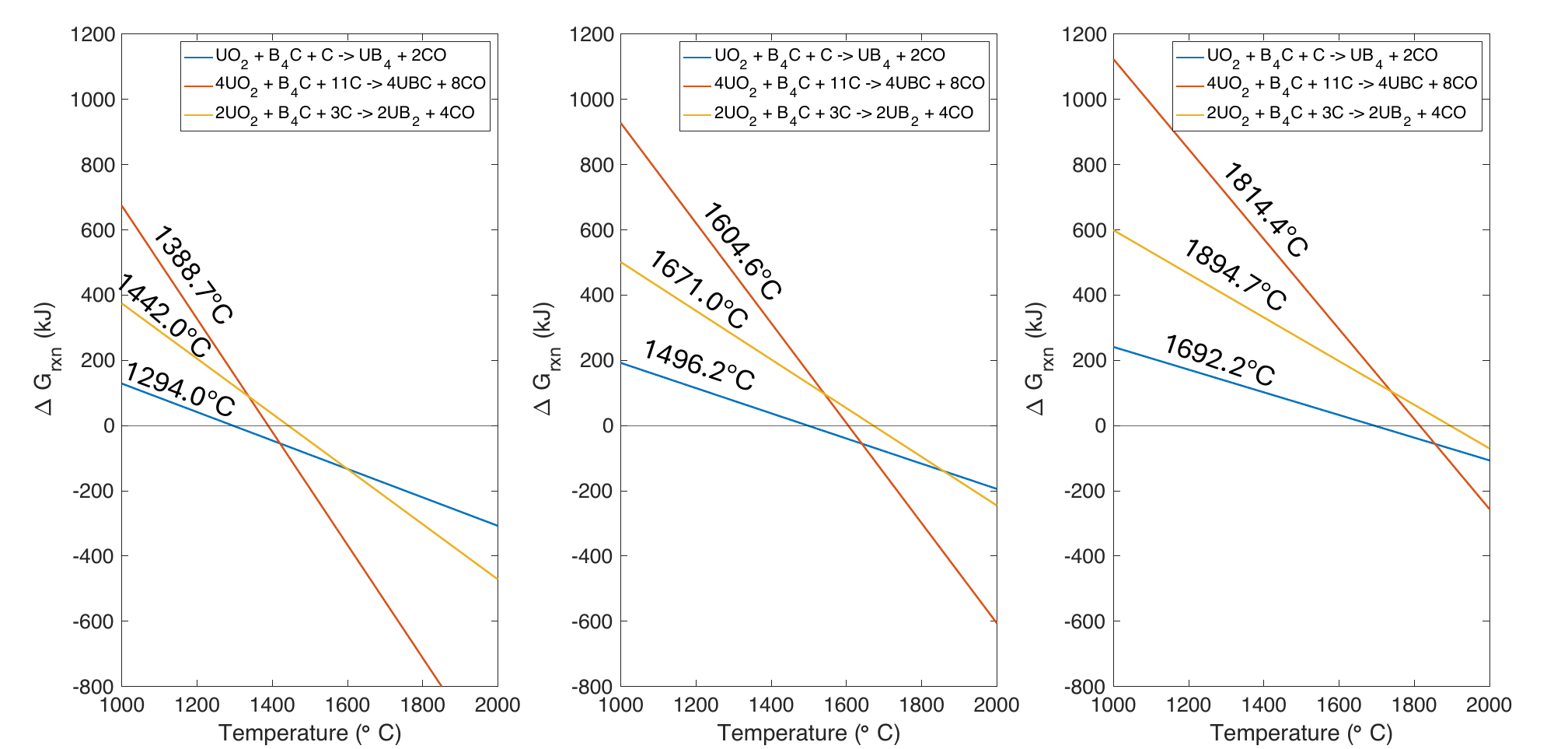}
    \caption{
    Temperature dependence of the Gibbs free energy of formation ($\Delta G_{rxn}$) for the reactions leading to \ce{UB4}, \ce{UBC}, and \ce{UB2} formation at varying (a) low – \qty{50}{\Pa}, (b) intermediate – \qty{10}{\kPa}, and (c) high – \qty{101}{\kPa} \ce{CO} partial pressures.}
    \label{fig:Figure2}
\end{figure}

The Gibbs free energy of the reaction was calculated using Equations (6) and (7) compared with reaction for UB\textsubscript{2 }synthesis as shown in Figure \ref{fig:Figure2} for three different CO partial pressures of CO gas. Meanwhile, since there is only sparse experimental data on the thermophysical properties of \ce{UBC}, a simple linear fit for the Gibbs free energy of formation was performed to calculate \(\Delta\)G\textsubscript{rxn } using the values  \(\Delta\)G\textsubscript{f } at 25\textsuperscript{o}C and 1600\textsuperscript{o}C from Rogl \textit{et al.} \autocite{ROGL198974}. The comparison in Figure \ref{fig:Figure2}, shows that, at CO partial pressures ranging from 50 Pa to 101 kPa, the relative spontaneity of these reactions with increasing temperature follows the order: T(\ce{UB4}) \(<\)T(\ce{UBC}) \(<\)T(\ce{UB2}). The thermodynamic stability maps which is constructed as a pressure dependent Ellingham-type plot for gas–solid or solid–solid reactions involving the release of gaseous species such CO gas shows the regions where the synthesis of \ce{UB4}, \ce{UB2} and \ce{UBC} is spontaneous (Figure S1). These thermodynamic results corroborate experimental observations indicating that the formation of \ce{UB4} and \ce{UBC} is favored under high CO partial pressures at 1700\textsuperscript{o}C, whereas the synthesis of \ce{UB2} is suppressed by the excess CO atmosphere. Each line represents a possible reaction pathway involving \ce{UO2}, \ce{B4C}, and C as starting feedstocks. The intersection points indicate the equilibrium temperatures where \(\Delta\)G\textsubscript{rxn}=0. With increasing CO partial pressure, all reactions become thermodynamically less favorable, shifting the equilibrium to higher temperatures.


\subsection{\ce{UB4} and \ce{UB4}–\ce{UBC} synthesis}

\color{blue}The XRD patterns of the starting \ce{UO2}, \ce{B4C}, and graphite feedstocks (Figure S2) confirm the phase purity of all materials, with all reflections matching their respective standard reference patterns and no detectable secondary phases. The SEM image of the \ce{UO2} powder is uniformly dispersed at the macroscale but exhibits significant microscale agglomeration, with fine particles forming clustered networks separated by inter-agglomerate porosity, as shown in Figure S3. To determine the choice of crucible for sintering pure \ce{UB4}, we used an alumina and a graphite crucible at 1500\textsuperscript{o}C. The XRD patterns of the synthesized samples is shown in Figure S4. The samples sintered in alumina crucibles show predominantly \ce{UB4} formation with minor \ce{UO2} phases, whereas samples synthesized in carbon crucibles exhibit mainly \ce{UO2} phase, indicating incomplete sintering at 1500\textsuperscript{o}C. The comparison also highlights the influence of sintering environments on phase evolution. \color{black}The sintering conditions for each \ce{UB4} sample and the corresponding phases from XRD analysis are summarized in Table~\ref{tab:UB4}. It is important to point out that the alumina sample holder contributes to the XRD signal which is denoted as \ce{Al2O3}* in Table~\ref{tab:UB4}. 


\begin{table}[htbp]
\caption{Processing conditions and resulting dominant phases for \ce{UB4} synthesis. \ce{Al2O3}* phases originate from the alumina sample holder.}
\label{tab:UB4}
\centering
\begin{tabular}{l c c c l}
\toprule
Stoichiometry & Ball-milling speed & Time & Temperature & Dominant phases \\
(\ce{UO2}: \ce{B4C}: C)& (rpm, time) & (h) & (°C) & (\ce{UB4}) \\
\midrule
(1.00, 1.04, 1.00) & 300, 4.5 h & 1 & 1500 & \ce{UB4}, \ce{Al2O3}*\\
(1.00, 1.04, 1.00) & 300, 4.5 h & 2 & 1500 & \ce{UB4}, \ce{Al2O3}*\\
(1.00, 1.04, 1.00) & 300, 4.5 h & 3 & 1500 & \ce{UB4} \\
(1.00, 1.04, 1.00) & 300, 4.5 h & 5 & 1500 & \ce{UB4} \\
(1.00, 1.04, 1.00) & 300, 4.5 h & 1 & 1450 & \ce{UB4}, \ce{Al2O3}*\\
(1.00, 1.04, 1.00) & 300, 4.5 h & 1 & 1550 & \ce{UB4}, \ce{Al2O3}*\\
\bottomrule
\end{tabular}
\end{table}



\begin{figure}[!htbp]
    \centering
    \includegraphics[width=1.0\linewidth]{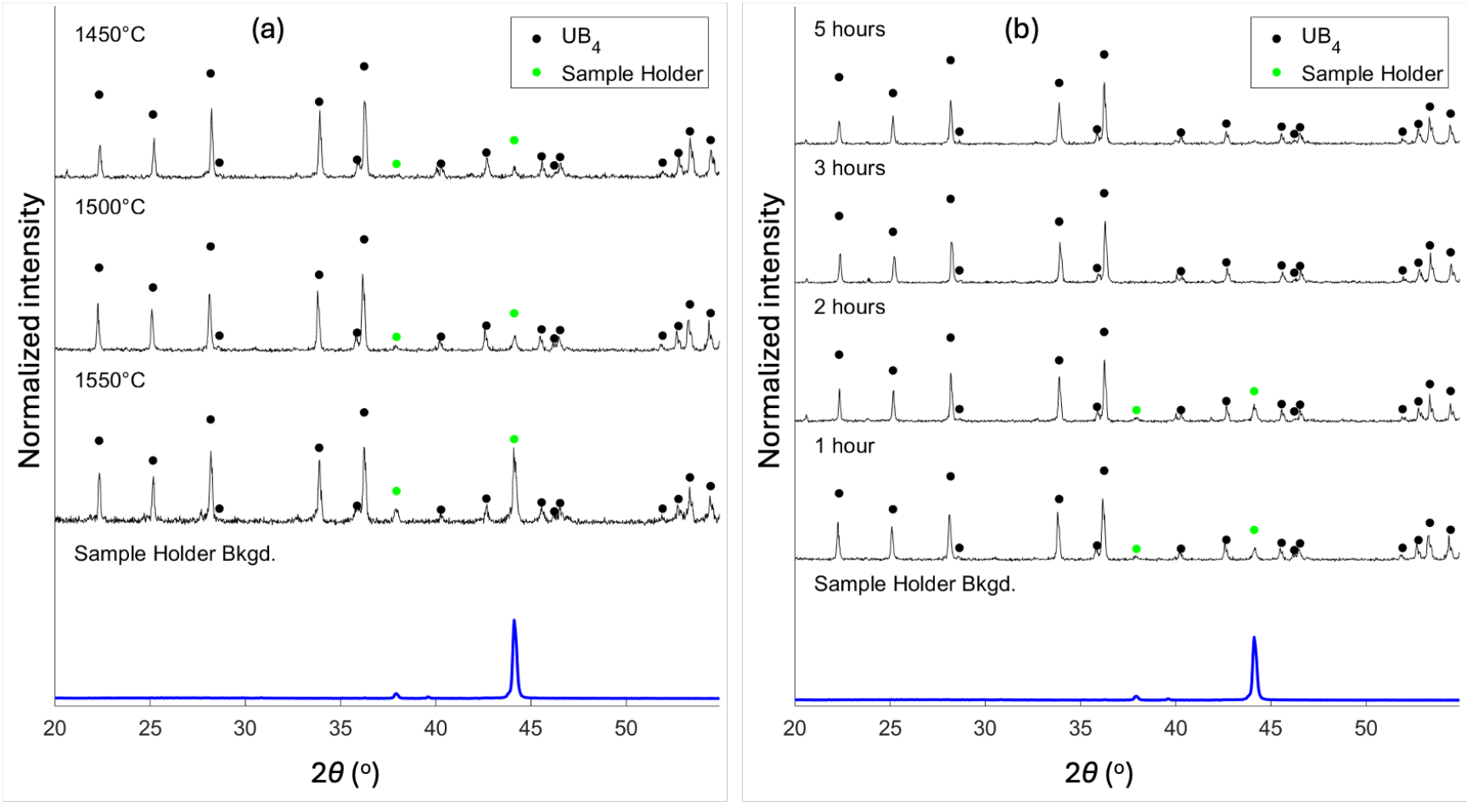}    \caption{
    (a) Effect of sintering temperature on the LXRD patterns of \ce{UB4} after \qty{1}{\hour} of sintering. 
    (b) Effect of sintering duration on the LXRD patterns of \ce{UB4} at a fixed temperature of \qty{1500}{\degreeCelsius}. 
    Peaks were indexed against PDF [01-085-4598] for \ce{UB4} taken from the International Centre of Diffraction Data (ICDD).
    }
    \label{fig:Figure3}
\end{figure}

The LXRD results shown in Figure \ref{fig:Figure3} include peaks corresponding to alumina, originating from the alumina sample holder. Initially, these were suspected to arise from a reaction between \ce{UB4} and the alumina crucible. However, background scans of the XRD sample holder confirmed that the aluminum peaks originated from the holder itself due to incomplete coverage by the sparse powder. By considering the contribution of the sample holder to the LXRD results, we found that all samples sintered at varied temperatures and durations show a dominant composition of \ce{UB4} phase. It is suspected that some phase fraction of \ce{UO2} is present within the samples but is not well resolved by the LXRD measurement. Furthermore, overlap between the high intensity \ce{UO2} (111) reflection and the UB\textsubscript{4 }(210) reflection may contribute to the obscured \ce{UO2} signal at this reflection. This suggests that \qty{1}{\hour} is sufficient for the synthesis of \ce{UB4} in small quantities at temperatures ranging from \qty{1450}{\degreeCelsius} to \qty{1500}{\degreeCelsius}.

The stoichiometry, sintering conditions, and resulting composition of the \ce{UB4}–\ce{UBC} composite are summarized in Table \ref{tab:stochio}. The initial stoichiometry resulted in UBC formation with predominant \ce{UO2} phases. Subsequent optimization of the starting stoichiometric feedstocks yielded predominantly UBC and \ce{UB4} phases with a minor \ce{UO2} contribution at 2 h and \qty{1700}{\degreeCelsius}. \color{blue}The indexed XRD patterns are shown in Figure S5 and S6. For further SXRD analysis we focused on the nearly equimolar \ce{UB4}–\ce{UBC} composite with $\approx 8 - 11$wt.\% \ce{UO2} phase fraction.\color{black}

\begin{table}[H]
\caption{Initial stoichiometry, sintering conditions, and resulting composition of the \ce{UB4} and \ce{UB4}--UBC samples synthesized via the carboborothermic  reduction route. Samples were sintered using a graphite crucible under flowing argon gas.}
\label{tab:stochio}
\centering
\begin{tabular}{l c c c l}
\toprule
Stoichiometry & Ball-milling speed & Time & Temperature & Dominant phases \\
 (\ce{UO2}: \ce{B4C}: C)& (rpm, time) & (\unit{\hour}) & (\unit{\degreeCelsius}) &(\ce{UB4} and UBC)\\\midrule
 (2.00, 1.04, 3.60)& 300, 4.5 h & 2& 1700&\ce{UB4}, UBC, 8-11 wt.\% \ce{UO2}\\
 (0.17, 0.12, 1.72)& 300, 2 h& 2& 1700&31 wt.\% \ce{UB4}, 65 wt.\% UBC, 4 wt.\% \ce{UO2}\\
 (0.30, 0.20, 3.00)& 300, 2 h& 2& 1700&50 wt.\% \ce{UB4}, 45 wt.\% UBC, \ce{B4C}\\
 (0.29, 0.20, 3.01)& 300, 2 h& 2& 1700&26 wt.\% \ce{UB4}, 66 wt.\% UBC, \ce{UO2} and \ce{B4C}\\ 
 (5.47, 0.58, 0.44)& 300, 2 h& 2& 1700&predominantly \ce{UB4}, \ce{UO2}\\ \bottomrule
\end{tabular}
\end{table}

\subsection{Microstructural Characterization}

\begin{figure}
    \centering
    \includegraphics[width=1.0\linewidth]{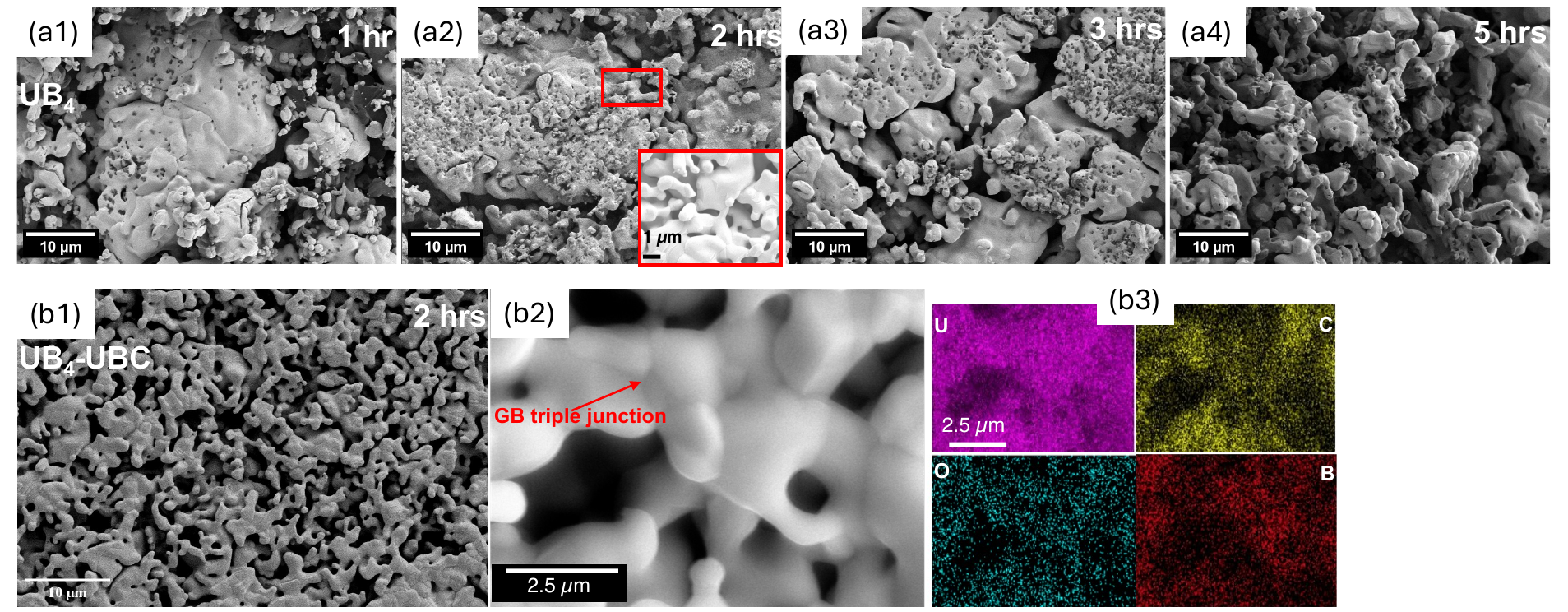}
    \caption{(a1-a4) SEM images of \ce{UB4} as a function of sintering time. The red region illustrate the coalescing grains during sintering. (b1–b3) SEM and EDX characterization of the \ce{UB4}–\ce{UBC} composite after 2 h of sintering. (b1) Low-magnification image shows a bicontinuous porous network with interconnected ligaments. (b2) Higher magnification reveals smooth necks and a grain boundary triple junction (arrow), indicative of active surface and grain boundary diffusion. (b3) Elemental maps demonstrate a homogeneous distribution of U, B, and C, confirming uniform phase distribution, while localized oxygen signals suggest minor surface oxidation. The combined morphological and compositional features indicate that sintering at this stage is dominated by non-densifying surface and grain boundary diffusion mechanisms.}
    \label{fig:Figure4}
\end{figure}

The microstructural evolution of \ce{UB4} as function of sintering time is shown in Figure \ref{fig:Figure4}\textbf(a1-a4) with initial particle coalescence leading to grain growth and densification. One hour of sintering shows region densification, a part of the \ce{UB4} sample also displays a highly porous morphology with partially bonded particles, which implies uneven solid-state diffusion. As the sintering duration increases from \qtyrange{2}{3}{\hour}, neck formation and grain coalescence become more pronounced, leading to a denser structure with improved interparticle bonding. 
However, after \qty{5}{\hour}, abnormal grain growth is observed, which points to prolonged exposure at high temperatures promoting atomic scale mobility, and resulting in grain coarsening and entrapment of localized porosity within larger grains. Meanwhile, the \ce{UB4}–\ce{UBC} composite sintered for \qty{1}{\hour} as shown in Figure \ref{fig:Figure4}b1 exhibits a noticeably finer and more homogeneous microstructure. 
The high-resolution SEM in Figure \ref{fig:Figure4}b2 shows faceted grains and well-defined grain-boundary triple junctions, which is indicative of efficient diffusion and grain growth \autocite{BIAN2024119602}. 
The presence of \ce{UBC} appears to inhibit grain growth by acting as a diffusion barrier and stabilizing the microstructure during sintering. This effect allows for higher relative density to be achieved in a shorter sintering time compared to monolithic \ce{UB4}.

Meanwhile, Figure \ref{fig:Figure4}b3 show that elemental mapping confirms the formation of a uniform \ce{UB4}–\ce{UBC} composite. Uranium and boron are evenly distributed, while there is preferential distribution of carbon at the grain boundaries, consistent with the presence of the \ce{UBC} phase. The minor oxygen signal compared to the U, C and B in the EDX map is indicative of the limited residual \ce{UB2} or surface oxidation. Overall, this shows that that excess carbon promotes formation of \ce{UBC} which enhances the sintering kinetics, refines the microstructure, and minimizes the oxidation of \ce{UB4}.

\section{Discussion}

\subsection{High temperature phase stability}

\begin{figure}
    \centering
    \includegraphics[width=1.0\linewidth]{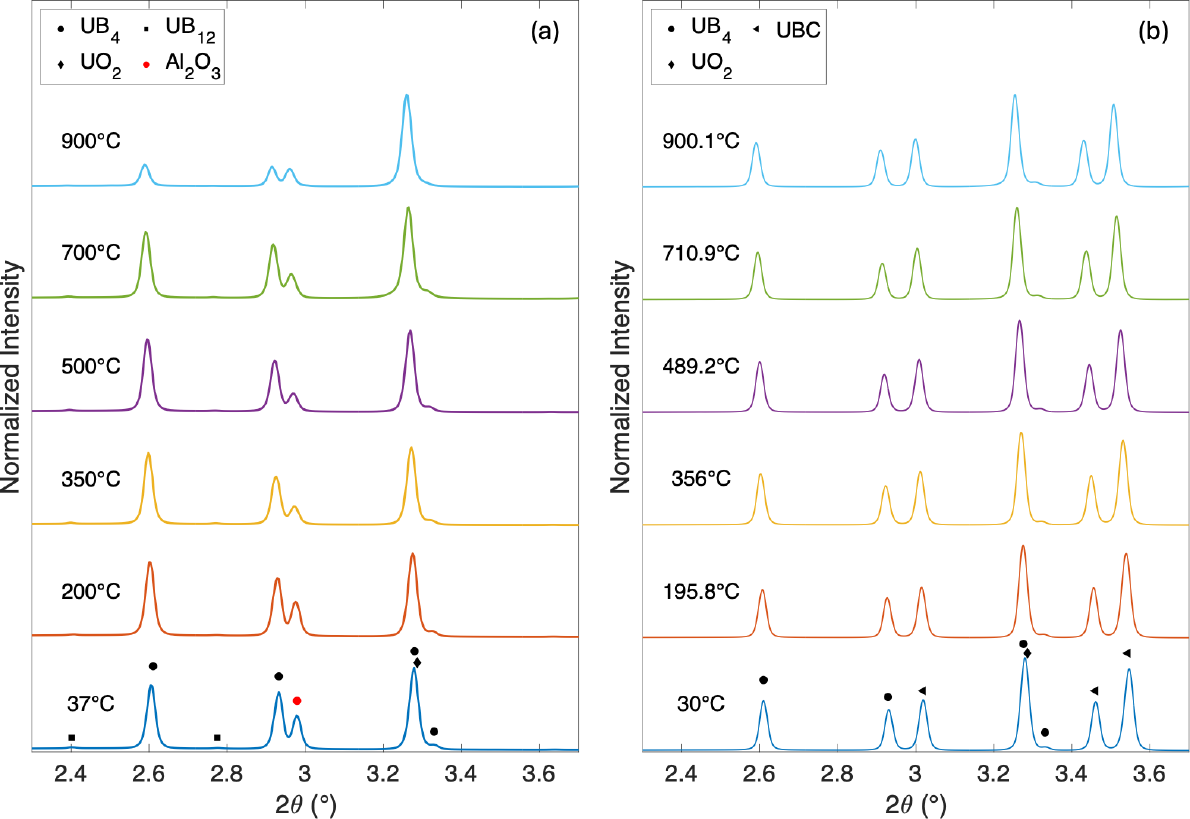}
    \caption{
    SXRD patterns showing the temperature-dependent phase evolution of (a) \ce{UB4} and (b) \ce{UB4}–\ce{UBC} during a step-by-step heating procedure in air.
    Diffraction patterns are displayed from room temperature to \qty{900}{\degreeCelsius}. 
    Peaks were indexed against PDF [01-085-4598] for \ce{UB4}, PDF [00-005-0550] for \ce{UO2}, PDF [01-083-2940] for \ce{UB12}, \ce{Al2O3} from Ishizawa et. al.\textsuperscript{35}, and PDF [04-008-8907] for \ce{UBC} (ICDD). }
    \label{fig:Figure5}
\end{figure}

The structural stability of the \ce{UB4} and \ce{UB4}–\ce{UBC} composite samples was investigated using the heating profiles shown in Figure \ref{fig:Figure1}. 
SXRD data was collected intermittently throughout the duration of the sample heating. 
A comparison of the SXRD results at intermittent temperatures can be observed in Figure \ref{fig:Figure5}. 
Phase fractions and lattice parameters were extracted through Rietveld refinement of the XRD data, and these values were subsequently used to calculate the volumetric thermal expansion of the \ce{UB4} and \ce{UBC} phases as shown in Figure \ref{fig:Figure5}. 

A notable phase fraction of \ce{Al2O3} is observed in the \ce{UB4} sample which comes from the alumina rod used for the \ce{UB4} sample mounting inside the capillaries. 
It is important to note that a \ce{UB12} and a \ce{UO2} phase were observed in the \ce{UB4} sample, highlighting the capability of high intensity, high energy synchrotron X-rays to resolve minor phase fractions in high atomic density samples. 
These phases were not observed in the LXRD measurements due to the low signal to noise ratio. Previous work also shows the presence of minor \ce{UB12} phase when synthesized with a 5 mol \% excess of boron carbide under the same thermal conditions \autocite{Guo2019} as used in this experiment. 
In Figure \ref{fig:Figure5}a, the diffraction peaks correspond to \ce{UB4}, \ce{UB12}, \ce{UO2}, and minor \ce{Al2O3} phases, with \ce{UB4} peaks decreasing and \ce{UO2} peaks becoming more pronounced as temperature increases, indicating progressive oxidation. 
At \qty{900}{\degreeCelsius}, the pattern shows dominant oxide phases and diminished boride peaks. 
Meanwhile, in Figure \ref{fig:Figure5}b, \ce{UB4}–\ce{UBC} shows a similar oxidation trend, but the presence of \ce{UBC} seems to delay the onset of oxides, with \ce{UO2} peaks appearing at lower temperatures of \qty{\approx 195}{\degreeCelsius} and boride peaks persisting at higher temperatures compared to pure \ce{UB4}. 
The results suggest that the \ce{UB4}–\ce{UBC} composite oxidizes more gradually over the heating profile, with phase evolution occurring at slightly lower temperatures than in the pure \ce{UB4} sample. 
Even though the \ce{UB4}\textbf{–}\ce{UBC} samples were sealed, it is possible that water vapor out-gassing during curing of alumina glue may contribute to oxidation behavior \autocite{HighTempCeramicAdhesivesA2S1}.

\begin{figure}
    \centering
    \includegraphics[width=1.0\linewidth]{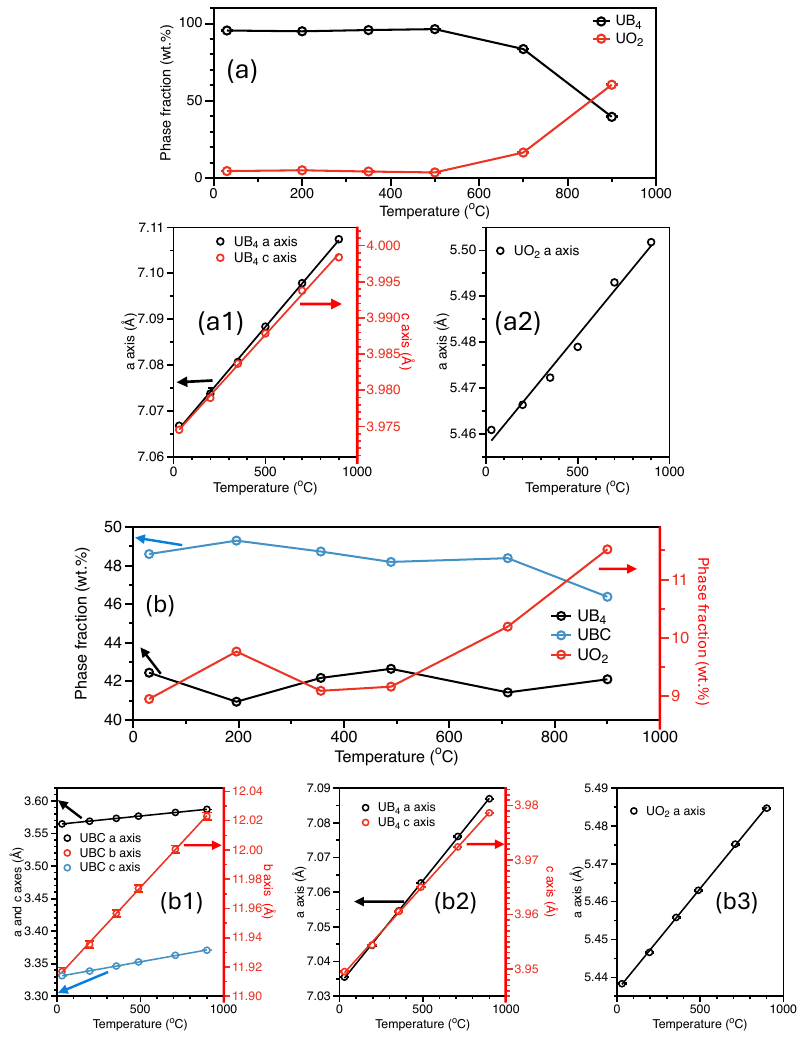}
    \caption{
    Temperature-dependent evolution of phase fractions and lattice parameters for pristine \ce{UB4} and \ce{UB4}–\ce{UBC} composite samples. 
    (a) Phase fractions of \ce{UB4} and \ce{UO2} as a function of temperature for the pristine \ce{UB4} sample. (a1–a2) Corresponding temperature dependence of the \ce{UB4} (\textit{a} and \textit{c}) and \ce{UO2} (\textit{a}) lattice parameters. 
    (b) Phase fractions of \ce{UB4}, \ce{UBC}, and \ce{UO2} for the \ce{UB4}–\ce{UBC} composite.
    (b1–b3) Temperature evolution of lattice parameters for \ce{UB4} (\textit{a} and \textit{c}), \ce{UBC} (\textit{a} and \textit{c}), and \ce{UO2} (\textit{a}), respectively.
    }
    \label{fig:Figure6}
\end{figure}

We have summarized the lattice parameters and phase fractions in Figure \ref{fig:Figure6}. 
At room temperature, the lattice parameters for \ce{UB4}–\ce{UBC} composite are similar to those of the pure \ce{UB4} phase. 
These results are similar to measurements by Bertaut \textit{et al}. where \textit{a} is \qty{7.080}{\angstrom}, and \textit{c} is \qty{3.978}{\angstrom} \autocite{Blum:a01065} while Zalkin \textit{et al.} reported \textit{a} equal to \qty{7.075}{\angstrom} and \textit{c} equal to \qty{3.979}{\angstrom} \autocite{Zalkin:a00862}. 
Additionally, the room temperature lattice parameter of the \ce{UBC} phase is comparable to reported values by Rogl \textit{et al.} where \textit{a} is \qty{3.5927}{\angstrom}, \textit{b} is \qty{11.9889}{\angstrom}, and \textit{c} is \qty{3.3474}{\angstrom} \autocite{Rogl1990ActinoidmetalBoronCarbides}. 
Furthermore, the Rietveld refinement of the \ce{UB4}–\ce{UBC} composite indicates phase fractions of \qty{42}{\weightfraction} \ce{UB4}, \qty{48}{\weightfraction} \ce{UBC}, and \qty{10}{\weightfraction} \ce{UO2}, respectively. 
These proportions remain stable throughout the heating process, demonstrating the thermal stability of the composite. 
At high temperatures, there is a slight increase in the phase fraction of \ce{UO2} while there is a corresponding decrease in the \ce{UB4} phase. 
For instance, in the pristine \ce{UB4} sample, at a critical temperature of about \qty{700}{\degreeCelsius}, a rapid increase of the \ce{UO2} phase is observed. 
This result would suggest that the \ce{UB4} sample experienced high temperature oxidation during heating. 

We calculated the coefficient of thermal expansion for the \ce{UB4} phase in the pristine \ce{UB4} sample, and for the \ce{UB4} and \ce{UBC} phases in the \ce{UB4}–\ce{UBC} composite to understand the extent of thermal strain induced during heating. 
The CTE of the \ce{UB4} and \ce{UBC} phases for each lattice direction are summarized in Table \ref{tab:cte} and compared with values from previous measurements. 
The measured directionally averaged linear coefficient of expansion was \qty{\approx 30}{\percent} greater for the \ce{UB4} phase of the \ce{UB4}–\ce{UBC} composite compared with the \ce{UB4} phase of the pristine \ce{UB4} sample. 
Matterson \textit{et al}. reported a value of \qty{7.1e-6}{\kelvin^{-1}} for the \ce{UB4} phase measured by dilatometry \autocite{osti_4652114} which is lower than the value calculated here from SXRD data. 
For the \ce{UB4}–\ce{UBC} composite, a directionally averaged CTE of \qty{10.75e-6}{\kelvin^{-1}} was calculated for the \ce{UBC} phase.

\begin{table}[htbp]
\caption{Mean linear coefficients of thermal expansion for the \ce{UB4} phase of the pristine \ce{UB4} and the \ce{UB4} and UBC phases of the \ce{UB4}--UBC composite compared to previous studies.}
\label{tab:cte}
\centering
\begin{tabular}{@{} ll c c c c l @{}}
\toprule
Sample & Phase & \makebox[0pt][c]{a-axis} & b-axis & c-axis & Averaged & Measurement \\
 &  & \multicolumn{1}{c}{($\times 10^{-6}\,\mathrm{K}^{-1}$)} 
   & \multicolumn{1}{c}{($\times 10^{-6}\,\mathrm{K}^{-1}$)} 
   & \multicolumn{1}{c}{($\times 10^{-6}\,\mathrm{K}^{-1}$)} 
   & \multicolumn{1}{c}{($\times 10^{-6}\,\mathrm{K}^{-1}$)} 
   & technique \\
\midrule
Pristine \ce{UB4}      & \ce{UB4} & \(6.22(28)\)  & --         & \(7.57(47)\)  & 6.67  & Diffraction \\
\ce{UB4}--UBC          & UBC    & \(7.44(16)\)  & \(10.45(40)\) & \(14.35(52)\) & 10.75 & Diffraction \\
\ce{UB4}--UBC          & \ce{UB4} & \(8.43(24)\)  & --         & \(8.65(46)\)  & 8.50  & Diffraction \\
Pristine \ce{UB4}      & \ce{UB4} & --            & --         & --            & 7.1   & Dilatometry\autocite{osti_4652114} \\
Pristine \ce{UB4}      & \ce{UB4} & --            & --         & --            & 10.0  & Dilatometry\autocite{KARDOULAKI2020153216}  \\
Pristine \ce{UB2}      & \ce{UB2} & \(9.00\)      & --         & \(8.00\)      & 7.7   & Diffraction\autocite{Beckman1956} \\
Pristine \ce{UB2}      & \ce{UB2} & --            & --         & --            & 9.0   & Dilatometry\autocite{KARDOULAKI2020153216}  \\
\bottomrule
\end{tabular}
\end{table}

\subsection{\textit{In situ} high temperature oxidation behavior}

\subsubsection{Thermogravimetric analysis }

\begin{figure}
    \centering
    \includegraphics[width=1.0\linewidth]{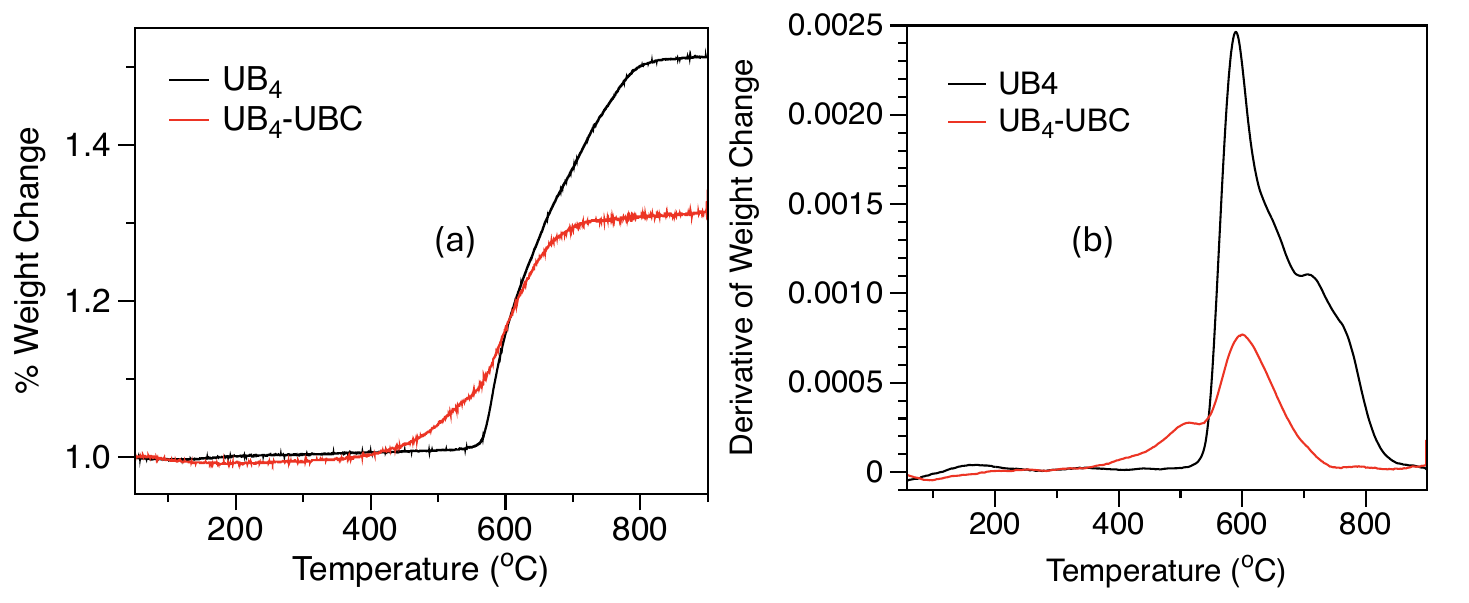}
    \caption{
    (a) Fractional mass change of \ce{UB4} and \ce{UB4}–\ce{UBC} with temperature under flowing dry air, measured by thermogravimetric analysis (TGA). 
    (b) The differential thermogravimetric (DTG) for \ce{UB4} and \ce{UB4}–\ce{UBC} under flowing dry air, measured by thermogravimetric analysis.
    }
    \label{fig:Figure7}
\end{figure}

The \ce{UB4} and the \ce{UB4}–\ce{UBC} composite were investigated via thermogravimetric analysis. 
The measurement and data analysis methodology for rendering fractional mass changes with evolving temperature are described in Section 2.2. 
The results for the fractional mass change under flowing dry air for the \ce{UB4} sample and the \ce{UB4}–\ce{UBC} composite sample are displayed in Figure \ref{fig:Figure7}a. 
The first derivative of the mass change for each sample is displayed in Figure \ref{fig:Figure7}b.

Figure \ref{fig:Figure7} displays the thermogravimetric (TGA) curves for pristine \ce{UB4} and the \ce{UB4}–\ce{UBC} composite under an oxidizing atmosphere. 
Pristine \ce{UB4} demonstrates negligible mass change from \qtyrange{\approx550}{580}{\degreeCelsius}, after which a sharp increase in weight occurs, resulting in a total mass gain of \qtyrange{\approx50}{52}{\percent} at \qty{900}{\degreeCelsius}. 
This pronounced weight gain signifies rapid oxidation at elevated temperatures and suggests the formation of \ce{U3O8} and \ce{B2O3} phases as shown in XRD phase analysis. This is consistent with Guo \textit{et al.} who observed the formation of \ce{U3O8} in the temperature range of \qtyrange{400}{600}{\degreeCelsius} and \ce{UB2O6} at \qtyrange{600}{800}{\degreeCelsius} \autocite{Guo2019}. 
In contrast, the \ce{UB4}–\ce{UBC} composite shows an earlier onset of weight gain around \qtyrange{450}{500}{\degreeCelsius}, but the mass increase proceeds more gradually compared to \ce{UB4}.
The total weight gain of the composite is within the range of \qtyrange{\approx30}{32}{\percent} at \qty{900}{\degreeCelsius}, \color{blue}which is consistent with theoretical weight increase of \qty{31}{\percent} \color{black}indicating a substantial reduction relative to pristine \ce{UB4}. These results indicate that the \ce{UBC} phase effectively suppresses high-temperature oxidation and limits oxygen uptake. The corresponding differential thermogravimetric (DTG) curves are presented in Figure \ref{fig:Figure7}b. 
Pristine \ce{UB4} displays a sharp and intense oxidation peak centered at \qtyrange{590}{620}{\degreeCelsius}, with a maximum derivative weight change of about \num{2.4e-3}, indicating a rapid oxidation rate. 
Additional shoulders at higher temperatures suggest multi-step oxidation processes. 
This corresponds to the formation of \ce{U3O8} and \ce{B2O3} in the first step followed by a second step where excess \ce{U3O8} react with \ce{B2O3} to form \ce{UB2O6} as follows:

\begin{equation}
3\,\mathrm{UB}_{4}\,(\mathrm{s}) + 13\,\mathrm{O}_{2}\,(\mathrm{g})
\rightarrow \mathrm{U}_{3}\mathrm{O}_{8}\,(\mathrm{s}) + 6\,\mathrm{B}_{2}\mathrm{O}_{3}\,(\mathrm{s})
\end{equation}

\begin{equation}
6\,\mathrm{B}_{2}\mathrm{O}_{3}\,(\mathrm{s}) + 2\,\mathrm{U}_{3}\mathrm{O}_{8}\,(\mathrm{s}) + \mathrm{O}_{2}\,(\mathrm{g})
\rightarrow 6\,\mathrm{UB}_{2}\mathrm{O}_{6}\,(\mathrm{s})
\end{equation}


Meanwhile, the \ce{UB4}–\ce{UBC} composite exhibits a broader and significantly lower-intensity peak, with a maximum of \num{\approx 0.75e-3}, confirming a markedly reduced oxidation rate and more controlled reaction kinetics. The results suggest that the incorporation of \ce{UBC} into \ce{UB4} leads to both a reduction in the oxidation rate and a decrease in total mass gain. 

\color{blue}The mechanism of oxidation of \ce{UB4} proceeds via a chemo-mechanical process initiated by rapid oxygen ingress through grain boundaries and surface defects. A strong thermodynamic drive favors the formation of uranium oxides \cite{Gueneau2002}. At elevated temperature and oxygen concentrations, \ce{U3O8} and \ce{UB2O6} are produced as shown in Figure \ref{fig:Figure8}a, enabling the reaction to proceed at moderate temperatures. Grain boundaries act as fast diffusion sites which facilitate oxidation between grains \cite{Clarke1987}, leading to localized expansion, particularly in the presence of \ce{U3O8}. This expansion induces stress, which promotes crack formation and fragmentation of the fuel. Concurrently, \ce{B2O3} may form a glassy phase that temporarily seals pores; however, at higher temperatures, it volatilizes, leaving voids. Consequently, the uranium and boron oxides layers transition from partial protection to a state where cracks and pores enhance oxygen transport, thereby accelerating oxidation.

In \ce{UB4}–\ce{UBC} composites, the oxidation mechanism may become more pronounced and spatially heterogeneous due to the interplay between phase boundaries and microstructural heterogeneity \cite{Chiang1997}. Interfaces between \ce{UB4} and UBC act as primary sites for oxygen ingress and stress concentration, resulting in earlier crack initiation compared to monolithic \ce{UB4}. When UBC forms a continuous and stable network, it may locally slow down oxygen diffusion. Nevertheless, differences in oxidation rates and thermal expansion between the phases generally cause interfacial separation, which facilitates crack propagation and increases oxygen uptake. As oxidation progresses, these differential reaction rates further elevate local stress, accelerating both fragmentation and the formation of an interconnected pore network. Ultimately, this leads to a highly porous fuel forms, mechanically unstable oxide layers dominated by \ce{U3O8} with either residual or depleted \ce{B2O3}. In the composites, this degradation may progress more rapidly due to the emergence of additional diffusion and failure pathways at interface boundaries.\color{black}


\subsubsection{\textit{In situ} phase stability of pre-oxidized \ce{UB4}-\ce{UBC}}

\begin{figure}
    \centering
    \includegraphics[width=1.0\linewidth]{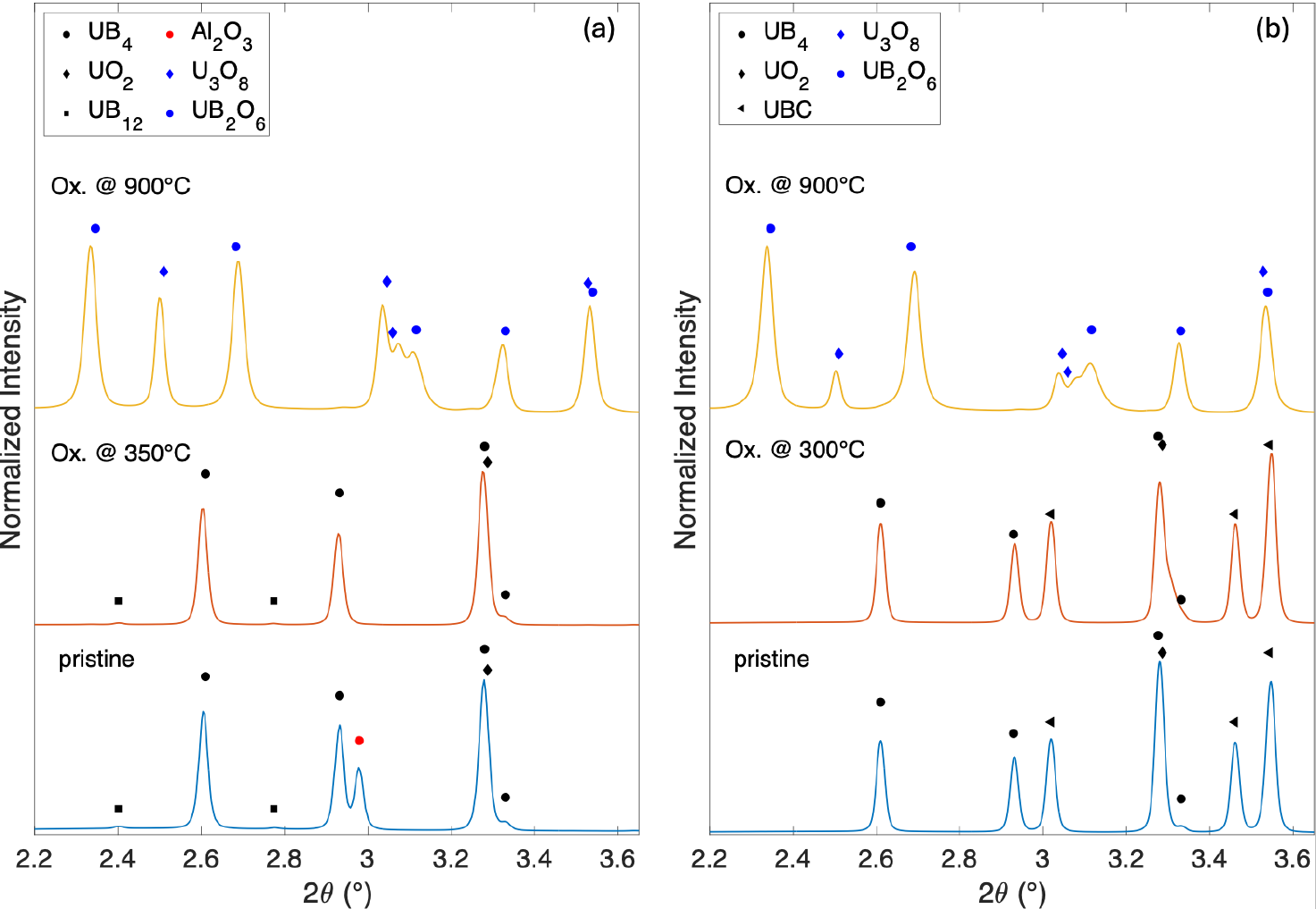}
    \caption{
    This figure shows XRD patterns comparing the oxidation behavior of (a) \ce{UB4} and (b) \ce{UB4}–\ce{UBC} composites under different atmospheric conditions (dry versus moist air) and temperatures.
    Peaks were indexed against PDF [01-085-4598] for \ce{UB4}, PDF [00-005-0550] for \ce{UO2}, PDF [01-083-2940] for \ce{UB12}, \ce{Al2O3} from Ishizawa \textit{et. al.}\cite{Ishizawa:a18575}, PDF [01-078-6745] for \ce{U3O8}, and PDF [04-012-6543] for \ce{UB2O6} taken from the ICDD.
    }
    \label{fig:Figure8}
\end{figure}

A microgram quantity of the synthesized \ce{UB4} and \ce{UB4}–\ce{UBC} was oxidized to \qty{300}{\degreeCelsius} / \qty{350}{\degreeCelsius} and \qty{900}{\degreeCelsius}, following the methodology described in Section 2.2. 
\textit{In situ} high-temperature SXRD measurements were performed to identify and index the phases present and to determine their corresponding crystal structures.
The resulting indexed peaks of the oxidized \ce{UB4} and \ce{UB4}–\ce{UBC} composite samples are compared in Figure \ref{fig:Figure8}. 
It should be noted again that the \ce{UB4} sample was oxidized in dry air and the \ce{UB4}–\ce{UBC} composite in moist air.

Oxidation studies revealed that both \ce{UB4} and \ce{UB4}–\ce{UBC} composite samples transformed predominantly into \ce{U3O8} and \ce{UB2O6} at \qty{900}{\degreeCelsius}, consistent with previously reported phases during high-temperature oxidation of \ce{UB4} \cite{Guo2019}. 
The results further indicate that oxidation of the \ce{UBC} phase proceeds through carbon volatilization, leading to the release of gaseous \ce{CO} and \ce{CO2} as shown in Equation 11 and 12:

\begin{equation}
2\,\mathrm{UBC}\,(\mathrm{s}) + 7\,\mathrm{O}_{2}\,(\mathrm{g})
\rightarrow 2\,\mathrm{UO}_{2}\,(\mathrm{s}) + \mathrm{B}_{2}\mathrm{O}_{3}\,(\mathrm{s}) + 2\,\mathrm{CO}_{2}\,(\mathrm{g})
\end{equation}
\begin{equation}
6\,\mathrm{UBC}\,(\mathrm{s}) + 19\,\mathrm{O}_{2}\,(\mathrm{g})
\rightarrow 2\,\mathrm{U}_{3}\mathrm{O}_{8}\,(\mathrm{s}) + 3\,\mathrm{B}_{2}\mathrm{O}_{3}\,(\mathrm{s}) + 6\,\mathrm{CO}_{2}\,(\mathrm{g})
\end{equation}

The onset of rapid oxidation was observed between \qtyrange{300}{350}{\degreeCelsius}, extending up to approximately \qty{900}{\degreeCelsius}, in agreement with the TGA data. 
Our results suggest that \ce{UB4} exhibits stability in oxidizing environments up to roughly \qty{350}{\degreeCelsius}, whereas the \ce{UB4}–\ce{UBC} composite remains stable only up to \qty{\approx 300}{\degreeCelsius}. 

\color{blue}This difference arises not only from compositional variations but also from the oxidation environment. The \ce{UB4}–\ce{UBC} samples were oxidized in moist air, which is known to accelerate oxidation kinetics compared to dry air. The presence of water vapor enhances oxygen transport and promotes the formation of volatile boron-containing species, leading to increased reaction rates and potential mass loss. As a result, oxidation in moist air can yield lower apparent mass gains and different phase evolution compared to dry air conditions, complicating direct comparison with \ce{UB4}oxidized in dry air.\color{black}


The lattice parameters of the pre-oxidized \ce{UB4} phase in dry air at \qty{350}{\degreeCelsius} are \textit{a} equal to \qty{7.070(72)}{\angstrom} and \textit{c} equal to \qty{3.975(10)}{\angstrom}. 
These lattice parameters exceed those of the pristine sample by less than \qty{\approx 1}{\percent}. Similarly, the relative phase \unit{\weightfraction} increased from \qty{4.5}{\percent} to \qty{5.5}{\percent} following the oxidation experiments. 
Meanwhile, the \ce{UBC}–UB\textsubscript{4 }composite observed a much greater increase in lattice volume following oxidation to \qty{300}{\degreeCelsius} in moist air. 
The lattice parameters of the \ce{UB4} phase were determined to be \textit{a} = \qty{7.075(3)}{\angstrom} and \textit{c} = \qty{3.973(3)}{\angstrom}, while those of the \ce{UBC} phase were found to be \textit{a} = \qty{3.586(3)}{\angstrom}, \textit{b} = \qty{11.986(3)}{\angstrom}, and \textit{c} = \qty{3.350(3)}{\angstrom}. 
There is a marginal increase in the relative phase fraction of \ce{UO2} from \qty{9}{\percent} to \qty{10}{\percent}. 
The samples oxidized to \qty{900}{\degreeCelsius} were refined for the resulting phase fractions and the \ce{UB2O6} lattice parameters. 
For the pristine \ce{UB4} sample oxidized to \qty{900}{\degreeCelsius} in dry air, the \ce{UB2O6} lattice parameters were determined such that \textit{a} is \qty{12.502(3)}{\angstrom}, \textit{b} is \qty{4.184(4)}{\angstrom}, and \textit{c} is \qty{10.490(2)}{\angstrom}. 
Similarly, for the \ce{UB4}–\ce{UBC} composite oxidized to \qty{900}{\degreeCelsius} in moist air, the lattice parameters were determined such that \textit{a} is \qty{12.430(4)}{\angstrom}, \textit{b} is \qty{4.162(4)}{\angstrom}, and \textit{c} is \qty{10.440(4)}{\angstrom} which is notably smaller than the parameters measured for the oxidized \ce{UB4} sample. 

\begin{figure}[H]
    \centering
    \includegraphics[width=1.0\linewidth]{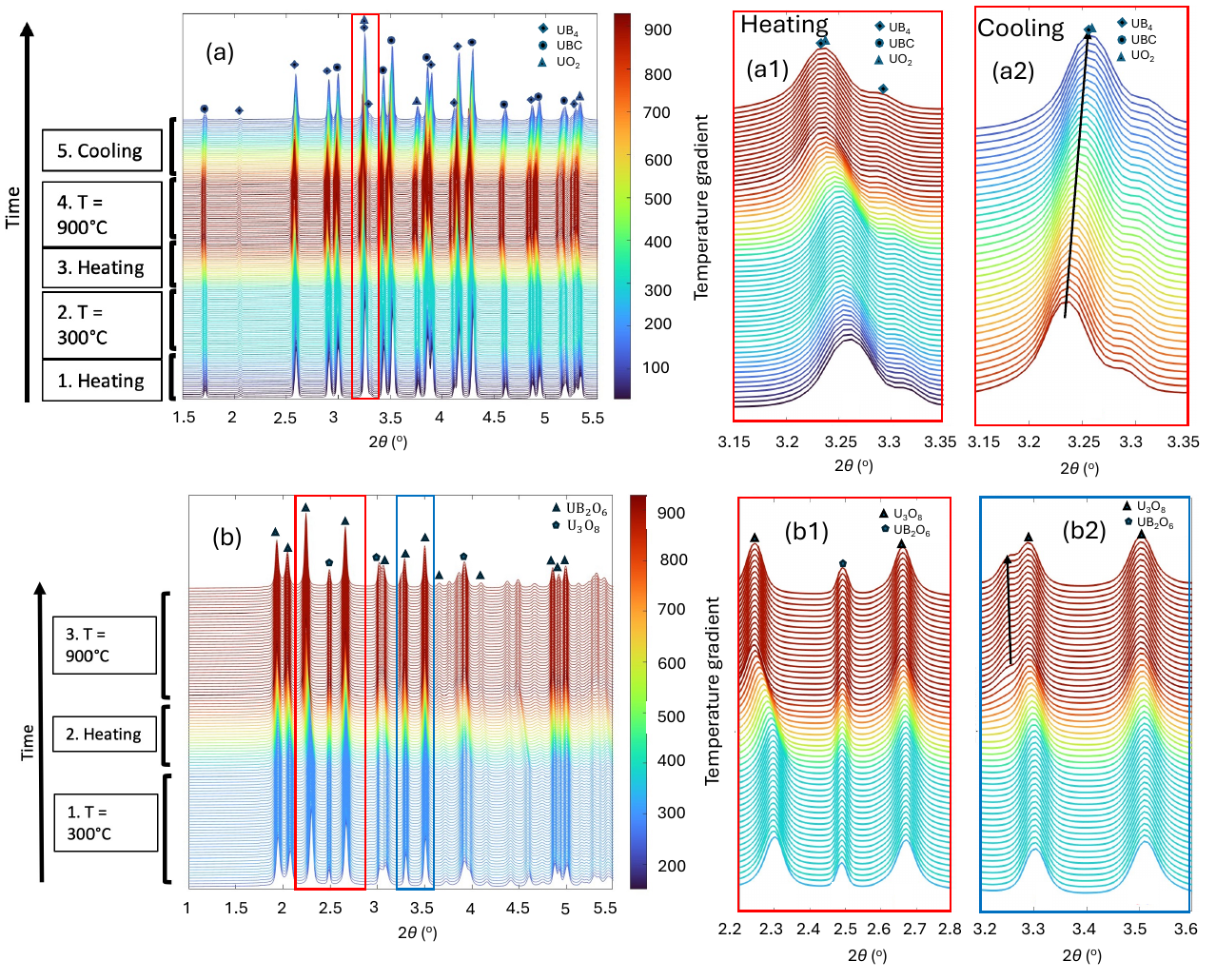}
    \caption{
    \textit{In situ} SXRD patterns showing the thermal stability of pristine and pre-oxidized of \ce{UB4}–\ce{UBC}. 
    (a) Contour plots illustrate diffraction intensity as a function of temperature and $2\theta$, highlighting the sequential phase transitions during heating and cooling. 
    Enlarged regions show the formation and evolution of \ce{UO2}, \ce{UBC}, and \ce{UB4} during (a1) heating (a2) cooling. 
    (b) XRD as a function of temperature highlighting the sequential phase transitions during heating. 
    The heating shows the persistence of oxide phases (\ce{UB2O6}, \ce{U3O8}) after high-temperature oxidation. 
    (b1) and (b2) The intensity of the XRD peak as a function of temperature shows the progressive oxidation and boron loss leading to \ce{U3O8} and \ce{UB2O6}  formation at elevated temperatures. 
    }
    \label{fig:Figure9}
\end{figure}

To further understand the high-temperature thermal stability of the \ce{UB4}–\ce{UBC} phase, we heated the sample at a rate of \qty{10}{\degreeCelsius\per\minute}.
We observed a minor \ce{UO2} phase that persisted throughout the heating and cooling process, as shown in Figure \ref{fig:Figure9}.
The overlapping \ce{UB4} and \ce{UO2} phases show a diffraction peak shift toward lower $2\theta$ (Figure \ref{fig:Figure9}a1), corresponding to lattice expansion as the temperature increases. 
This suggest that there is a thermal dilation of the \ce{UB4}, \ce{UBC}, and \ce{UO2} phases. 
The extent of the shift indicates that the phases have different thermal expansion coefficients, as shown in Table \ref{tab:oxidation_onset}. 
Furthermore, during the cooling stage, the diffraction peak shifted towards higher $2\theta$ values, which means that there is lattice contraction, as shown in Figure \ref{fig:Figure9}a2. 
The asymmetric shifting of the peaks may be due to residual strain, microstructural distortion, or irreversible phase evolution resulting from oxidation or partial phase decomposition \autocite{ROUSSEAU200610, ZALKIND2017202}.

The resulting relative weight percentages of the dominant phases \ce{UB2O6} and \ce{U3O8} in the samples oxidized to \qty{900}{\degreeCelsius} are presented in Figure S7. 
By nature of the different initial phase composition and the differing oxidation conditions, the phase fraction of \ce{U3O8} is \qty{\approx 15}{\percent} greater in the \ce{UB4} sample as compared to the \ce{UB4}–\ce{UBC} sample. 
Considering the reaction products and the oxidation peaks in the TGA measurements, the reaction of \ce{UB4}–\ce{UBC} is likely to have proceeded at a lower temperature than that of the \ce{UB4} composite. 
The pre-oxidized sample predominantly contains \ce{UB2O6} and \ce{U3O8}, which increase gradually, as shown by the peaks in Figure \ref{fig:Figure9}b. 
This implies progressive oxidation of the uranium boride phases. 
A further closer look at the diffraction patterns shown in Figure \ref{fig:Figure9}b1 – b2, is consistent with similar behavior during the oxidation of \ce{UB2} in flowing synthetic air \autocite{MISTARIHI2023154417}. 
In our case, our sample was sealed with epoxy which expels oxygen at high temperature, serving as the precursor for oxidation. The systematic increase in intensity of the oxide peaks reflects crystallization and growth of these phases during oxidation.

While the ATFs concepts have promising high thermal conductivity and high uranium loading, high temperature oxidation in steam and air is still a concern. The oxidation of uranium nitride \autocite{Jolkkonen04052017}, uranium carbide \autocite{GASPARRINI2024154944}, uranium diboride, and its composite \autocite{TURNER2020151919, MISTARIHI2023154417, TURNER2022153471} have been investigated using TGA and differential scanning calorimetry analysis. The oxidation behavior of uranium borides, carbides, silicides, and nitrides have been investigated under air and steam atmospheres, as summarized in Table \ref{tab:oxidation_onset}. 
The onset of oxidation depends strongly on composition and oxidation environments. 
The UC starts to oxidize from \qty{264}{\degreeCelsius} in air \autocite{GONCHAROV2023154446} while the oxidation of \ce{UB2} is close to \qty{\approx 700}{\degreeCelsius} in synthetic air \autocite{MISTARIHI2023154417}. 
Meanwhile, \ce{UB4}–\ce{UBC} composite exhibited the lower oxidation onset compared to \ce{UB2} and \ce{UB4}, which is due to the multiphase interfaces that may lead to rapid pulverization of the \ce{B2O3} formed in the early stage of oxidation.

\begin{table}[htbp]
\caption{Onset of oxidation temperature, oxidation environments, phase composition, and uranium loading of \ce{UB4}, \ce{UB4}--UBC, \ce{UB2}, UN, UC and other uranium borides-containing composites.}
\label{tab:oxidation_onset}
\centering
\begin{tabularx}{\textwidth}{@{} l c c X c @{}}
\toprule
\makecell[l]{Sample} 
 & \makecell{Onset of oxidation\\ temperature} 
 & \makecell{Oxidation\\ environment} 
 & \makecell[l]{Oxidation products} 
 & \makecell{Uranium loading\\ (\unit{\g\per\cm^3})} \\
\midrule
\ce{UB4} 
 & \qty{550}{\degreeCelsius} 
 & Synthetic air 
 & \makecell[l]{\ce{UB2O6}, \ce{U3O8},\\ \ce{UO2}} 
 & 7.94 \\

\ce{UB4}--UBC 
 & \qty{400}{\degreeCelsius} 
 & Ambient air 
 & \ce{UB2O6}, \ce{U3O8} 
 & $\sim$9.5 \\

\ce{UB2}\autocite{MISTARIHI2023154417} 
 & \qtyrange{534}{692}{\degreeCelsius} 
 & Synthetic air 
 & \makecell[l]{\ce{UB2O6}, \ce{U3O8},\\ \ce{UO2}} 
 & 11.7 \\

UC\autocite{GASPARRINI2024154944} 
 & \qty{264}{\degreeCelsius} 
 & Synthetic air 
 & \makecell[l]{\ce{U2C3}, \ce{UO2},\\ \ce{U3O8}} 
 & 12.98 \\

\makecell[l]{50 wt\% \ce{UB2}--50 wt\%\\\ce{U3Si2}\autocite{TURNER2020151919}} 
 &\qtyrange{548}{ 575}{\degreeCelsius} 
 & Steam 
 & \makecell[l]{\ce{UO2}, \ce{U3Si2H2}} 
 & -- \\

\makecell[l]{10 wt\% \ce{UB2}--90 wt\%\\\ce{U3Si2}\autocite{TURNER2020151919}} 
 & \qty{553}{\degreeCelsius} 
 & Steam 
 & \makecell[l]{\ce{UO2}, \ce{U3Si2H2}} 
 & -- \\

\makecell[l]{UN--5, 25,\\ 50 wt\% \ce{UB2}\autocite{TURNER2022153471}} 
 & \qtyrange{500}{600}{\degreeCelsius} 
 & Steam 
 & \makecell[l]{\ce{U2N3}, \ce{UO2}, \ce{B2O3}} 
 & -- \\

UN\autocite{Jolkkonen04052017} 
 & \qtyrange{400}{500}{\degreeCelsius} 
 & Steam 
 & \ce{UO2}, \ce{U3O8} 
 & 13.51 \\
\bottomrule
\end{tabularx}
\end{table}

\section{Conclusions}

In this work, we synthesized \ce{UB4} and \ce{UB4}–\ce{UBC} composite via an industrially scalable carboborothermic  reduction of a \ce{UO2}–\ce{B4C}–\ce{C} feedstock at temperatures between \qtyrange{1450}{1700}{\degreeCelsius}. 
A systematic optimization of sintering conditions is evaluated for \ce{UB4} and \ce{UB4}–\ce{UBC} synthesis. The XRD results indicate that high-purity \ce{UB4} can be produced at \qtyrange{1450}{1550}{\degreeCelsius} with sintering durations of \qtyrange{1}{5}{\hour}, when employing a stoichiometric excess \ce{B4C} feedstock.
Similarly, synthesis at \qty{1700}{\degreeCelsius} with excess graphite and \ce{UO2} resulted in the formation of a \ce{UB4}–\ce{UBC} composite. The \ce{UB4}–\ce{UBC} has a higher uranium loading than monolithic \ce{UB4}, comparable to \ce{UO2}. 
The \ce{UB4} sample synthesized at \qty{1500}{\degreeCelsius} for \qty{1}{\hour}, and the \ce{UB4}–\ce{UBC} composite (\ce{UB4}: \ce{UBC} $\approx 4:5$) containing a minor \ce{UO2} impurity phase, were further characterized by \textit{in situ} high temperature SXRD analysis from room temperature to \qty{900}{\degreeCelsius}. 
The room temperature lattice parameters and linear thermal expansion coefficients were determined for the relevant phases. 
Notable differences in the \ce{UB4} phase expansion coefficient were observed between the two samples, which is due to anisotropy in the \textit{c}/\textit{a} ratio.

The oxidation behavior of the composites was investigated using TGA and subsequent post-oxidation structural characterization. 
The TGA results indicated oxidation onset temperatures of approximately \qty{400}{\degreeCelsius} for the \ce{UB4}–\ce{UBC} composite and \qty{550}{\degreeCelsius} for the \ce{UB4} sample. 
\color{blue}Although oxidation in \ce{UB4}–\ce{UBC} begins earlier than in \ce{UB4}, its progression is slower due to the presence of the \ce{UBC} phase. This is evident from the phase fractions of \ce{U3O8} and \ce{UB2O6} observed in the \textit{in situ} SXRD analysis, despite preoxidation in moist air, which would typically accelerate oxidation kinetics compared to the dry-air conditions used for \ce{UB4}. \color{black} \ce{UB4}–\ce{UBC} and \ce{UB4} display desirable attributes for an accident tolerant nuclear fuel, with promising oxidation behavior highlighted in this study. 
While this study provided insight into the synthesis, structural stability, and oxidation behavior of a novel \ce{UB4}–\ce{UBC} composite and closed the knowledge gap regarding \ce{UB4}, future studies should focus on further \textit{in situ} study of \ce{UB4} and \ce{UB4}–\ce{UBC} oxidation behavior under accident scenarios to gain an understanding of the oxidation reaction kinetics. 

\section{Acknowledgement}

This work was funded by the Faculty Startup Fund and Research Support Committee grant from the Massachusetts Institute of Technology. E.J. acknowledges support from the John Hardwick Career Development Chair and R.M. acknowledges the Theos J. Thompson memorial fellowship. This research used beamline 28-ID-2 (XPD) of the National Synchrotron Light Source II, a US DOE Office of Science user facility operated for the DOE Office of Science by Brookhaven National Laboratory under contract number DE-SC0012704; This research utilized resources at XPD beamline of NSLS-II that were partially supported by the U.S. Department of Energy, Office of Nuclear Energy under DOE Idaho Operations Office Contract DE-AC07-051D14517 as part of Nuclear Science User Facilities experiment (\#24-4876). 

\appendix

\clearpage

\printbibliography

@book{Chiang1997,
  author    = {Yet-Ming Chiang and Dunbar P. Birnie III and W. David Kingery},
  title     = {Physical Ceramics: Principles for Ceramic Science and Engineering},
  publisher = {Wiley},
  year      = {1997}
}

@article{Clarke1987,
author  = {David R. Clarke},
title   = {On the Equilibrium Thickness of Intergranular Glass Phases in Ceramic Materials},
journal = {Journal of the American Ceramic Society},
volume  = {70},
number  = {1},
pages   = {15--22},
year    = {1987},
doi     = {10.1111/j.1151-2916.1987.tb04846.x},
url     = {https://ceramics.onlinelibrary.wiley.com/doi/10.1111/j.1151-2916.1987.tb04846.x}
}

@article{Gueneau2002,
  author  = {C. Gu{\'e}neau and M. Baichi and D. Labroche and C. Chatillon and B. Sundman},
  title   = {Thermodynamic Assessment of the Uranium--Oxygen System},
  journal = {Journal of Nuclear Materials},
  volume  = {304},
  number  = {2},
  pages   = {161--175},
  year    = {2002},
  issn    = {0022-3115},
  doi     = {10.1016/S0022-3115(02)00878-4},
  url     = {https://www.sciencedirect.com/science/article/pii/S0022311502008784}
}

@BOOK{Carmack2013-bs,
  title  = "Overview of the {U.S}. {DOE} Accident Tolerant Fuel Development
            Program",
  author = "Carmack, Jon and {Frank} and Bragg-Sitton, Shannon M and Lance, L",
  year   =  2013
}

@article{JOSSOU201741,
title = {A first principles study of the electronic structure, elastic and thermal properties of UB2},
journal = {Journal of Nuclear Materials},
volume = {490},
pages = {41-48},
year = {2017},
issn = {0022-3115},
doi = {https://doi.org/10.1016/j.jnucmat.2017.04.006},
url = {https://www.sciencedirect.com/science/article/pii/S0022311516309771},
author = {Ericmoore Jossou and Linu Malakkal and Barbara Szpunar and Dotun Oladimeji and Jerzy A. Szpunar},
}

@book{IAEA2008Thermophysical,
  author       = {{International Atomic Energy Agency}},
  title        = {Thermophysical Properties of Materials for Nuclear Engineering: A Tutorial and Collection of Data},
  institution  = {International Atomic Energy Agency},
  address      = {Vienna, Austria},
  year         = {2008},
  note         = {Nuclear Power Technology Development Section}
}

@article{KARDOULAKI2020153216,
title = {Thermophysical and mechanical property assessment of UB2 and UB4 sintered via spark plasma sintering},
journal = {Journal of Alloys and Compounds},
volume = {818},
pages = {153216},
year = {2020},
issn = {0925-8388},
doi = {https://doi.org/10.1016/j.jallcom.2019.153216},
url = {https://www.sciencedirect.com/science/article/pii/S0925838819344627},
author = {E. Kardoulaki and J.T. White and D.D. Byler and D.M. Frazer and A.P. Shivprasad and T.A. Saleh and B. Gong and T. Yao and J. Lian and K.J. McClellan},
}

@article{GONZALES2021153026,
title = {Challenges and opportunities to alloyed and composite fuel architectures to mitigate high uranium density fuel oxidation: uranium silicide},
journal = {Journal of Nuclear Materials},
volume = {553},
pages = {153026},
year = {2021},
issn = {0022-3115},
doi = {https://doi.org/10.1016/j.jnucmat.2021.153026},
url = {https://www.sciencedirect.com/science/article/pii/S002231152100249X},
author = {Adrian Gonzales and Jennifer K. Watkins and Adrian R. Wagner and Brian J. Jaques and Elizabeth S. Sooby},
}

@article{TERRANI2020152267,
title = {Accelerating nuclear fuel development and qualification: Modeling and simulation integrated with separate-effects testing},
journal = {Journal of Nuclear Materials},
volume = {539},
pages = {152267},
year = {2020},
issn = {0022-3115},
doi = {https://doi.org/10.1016/j.jnucmat.2020.152267},
url = {https://www.sciencedirect.com/science/article/pii/S0022311519316733},
author = {Kurt A. Terrani and Nathan A. Capps and Matthew J. Kerr and Christina A. Back and Andrew T. Nelson and Brian D. Wirth and Steven L. Hayes and Chris R. Stanek},
}

@techreport{Faibish2021AFQ,
  author       = {Faibish, R.},
  title        = {Accelerated Fuel Qualification White Paper},
  institution  = {Accelerated Fuel Qualification Working Group White Paper Task Force},
  year         = {2021}
}

@article{Guo2019,
author = {Guo, Hangxu and Wang, Jieru and Chen, Denglei and Tian, Wei and Cao, Shiwei and Chen, Desheng and Tan, Cunmin and Deng, Qihuang and Qin, Zhi},
title = {Boro/carbothermal reduction synthesis of uranium tetraboride and its oxidation behavior in dry air},
journal = {Journal of the American Ceramic Society},
volume = {102},
number = {3},
pages = {1049-1056},
doi = {https://doi.org/10.1111/jace.15987},
url = {https://ceramics.onlinelibrary.wiley.com/doi/abs/10.1111/jace.15987},
eprint = {https://ceramics.onlinelibrary.wiley.com/doi/pdf/10.1111/jace.15987},
year = {2019}
}

@article{TURNER2020152388,
title = {Synthesis of candidate advanced technology fuel: Uranium diboride (UB2) via carbo/borothermic reduction of UO2},
journal = {Journal of Nuclear Materials},
volume = {540},
pages = {152388},
year = {2020},
issn = {0022-3115},
doi = {https://doi.org/10.1016/j.jnucmat.2020.152388},
url = {https://www.sciencedirect.com/science/article/pii/S002231152030996X},
author = {J. Turner and F. Martini and J. Buckley and G. Phillips and S.C. Middleburgh and T.J. Abram},
}

@article{BURR201945,
title = {Defect evolution in burnable absorber candidate material: Uranium diboride, UB2},
journal = {Journal of Nuclear Materials},
volume = {513},
pages = {45-55},
year = {2019},
issn = {0022-3115},
doi = {https://doi.org/10.1016/j.jnucmat.2018.10.039},
url = {https://www.sciencedirect.com/science/article/pii/S0022311518311000},
author = {P.A. Burr and E. Kardoulaki and R. Holmes and S.C. Middleburgh},
}

@article{Zalkin:a00862,
author = "Zalkin, A. and Templeton, D. H.",
title = "{The crystal structures of CeB${\sb 4}$ ThB${\sb 4}$ and UB${\sb 4}$}",
journal = "Acta Crystallographica",
year = "1953",
volume = "6",
number = "3",
pages = "269--272",
month = "3",
doi = {10.1107/S0365110X53000764},
url = {https://doi.org/10.1107/S0365110X53000764},
}

@article{osti_4652114,
  author       = {Matterson, K J and Jones, H},
  title        = {A STUDY OF THE TETRABORIDES OF URANIUM AND THORIUM},
  url          = {https://www.osti.gov/biblio/4652114},
  journal      = {Trans. Brit. Ceram. Soc.},
  volume       = {Vol: 60},
  place        = {Country unknown/Code not available},
  year         = {1961},
  month        = {07}}

@article{MENOVSKY1984519,
title = {The crystal growth of uranium tetraboride UB4 from the melt},
journal = {Journal of Crystal Growth},
volume = {70},
number = {1},
pages = {519-522},
year = {1984},
issn = {0022-0248},
doi = {https://doi.org/10.1016/0022-0248(84)90311-7},
url = {https://www.sciencedirect.com/science/article/pii/0022024884903117},
author = {A. Menovsky and J.J.M. Franse and J.C.P. Klaasse},
}

@article{ROGL198974,
title = {The ternary system uranium-boron-carbon},
journal = {Journal of Nuclear Materials},
volume = {165},
number = {1},
pages = {74-82},
year = {1989},
issn = {0022-3115},
doi = {https://doi.org/10.1016/0022-3115(89)90504-7},
url = {https://www.sciencedirect.com/science/article/pii/0022311589905047},
author = {Peter Rogl and Josef Bauer and Jean Debuigne},
}

@incollection{Rogl1990ActinoidmetalBoronCarbides,
  author    = {Rogl, P.},
  title     = {Actinoidmetal Boron Carbides},
  booktitle = {The Physics and Chemistry of Carbides, Nitrides and Borides},
  publisher = {Springer},
  year      = {1990},
  pages     = {269--277}
}

@article{KARDOULAKI2021152690,
title = {Fabrication and thermophysical properties of UO2-UB2 and UO2-UB4 composites sintered via spark plasma sintering},
journal = {Journal of Nuclear Materials},
volume = {544},
pages = {152690},
year = {2021},
issn = {0022-3115},
doi = {https://doi.org/10.1016/j.jnucmat.2020.152690},
url = {https://www.sciencedirect.com/science/article/pii/S0022311520312988},
author = {E. Kardoulaki and D.M. Frazer and J.T. White and U. Carvajal and A.T. Nelson and D.D. Byler and T.A. Saleh and B. Gong and T. Yao and J. Lian and K.J. McClellan},
}

@software{JADEPro,
  title        = {JADE Pro},
  author       = {{Materials Data, Inc.}},
  year         = {2023},
  note         = {X-ray diffraction analysis software},
  organization = {Materials Data, Inc.}
}

@article{Kabekkodu2024PDF5Plus,
  author  = {Kabekkodu, Soorya N. and Dosen, Anja and Blanton, Thomas N.},
  title   = {PDF-5+: A Comprehensive Powder Diffraction File for Materials Characterization},
  journal = {Powder Diffraction},
  volume  = {39},
  number  = {2},
  pages   = {47--59},
  year    = {2024},
  doi     = {10.1017/S0885715624000150}
}

@article{Toby:aj5212,
author = "Toby, Brian H. and Von Dreele, Robert B.",
title = "{{\it GSAS-II}: the genesis of a modern open-source all purpose crystallography software package}",
journal = "Journal of Applied Crystallography",
year = "2013",
volume = "46",
number = "2",
pages = "544--549",
month = "04",
doi = {10.1107/S0021889813003531},
url = {https://doi.org/10.1107/S0021889813003531},
}

@book{Barin1989ThermochemicalData,
  author    = {Barin, Ihsan and Platzki, Gregor},
  title     = {Thermochemical Data of Pure Substances},
  publisher = {VCH},
  address   = {Weinheim},
  year      = {1989},
  volume    = {304}
}

@article{BIAN2024119602,
title = {Coupling of alloy chemistry, diffusion and structure by grain boundary engineering in Ni–Cr–Fe},
journal = {Acta Materialia},
volume = {264},
pages = {119602},
year = {2024},
issn = {1359-6454},
doi = {https://doi.org/10.1016/j.actamat.2023.119602},
url = {https://www.sciencedirect.com/science/article/pii/S1359645423009308},
author = {Baixue Bian and Shabnam Taheriniya and G. Mohan Muralikrishna and Sandipan Sen and Christoph Gammer and Ingo Steinbach and Sergiy V. Divinski and Gerhard Wilde},
}

@misc{HighTempCeramicAdhesivesA2S1,
  title        = {High Temperature Ceramic Adhesives},
  howpublished = {Technical Bulletin A2-S1},
  note         = {Manufacturer technical bulletin}
}

@article{Blum:a01065,
author = "Blum, P. and Bertaut, F.",
title = "{Contribution {\`{a}} l'{\'{e}}tude des borures {\`{a}} teneur {\'{e}}lev{\'{e}}e en bore}",
journal = "Acta Crystallographica",
year = "1954",
volume = "7",
number = "1",
pages = "81--86",
month = "01",
doi = {10.1107/S0365110X54000151},
url = {https://doi.org/10.1107/S0365110X54000151},
}

@article{ROUSSEAU200610,
title = {A detailed study of UO2 to U3O8 oxidation phases and the associated rate-limiting steps},
journal = {Journal of Nuclear Materials},
volume = {355},
number = {1},
pages = {10-20},
year = {2006},
issn = {0022-3115},
doi = {https://doi.org/10.1016/j.jnucmat.2006.03.015},
url = {https://www.sciencedirect.com/science/article/pii/S0022311506002091},
author = {G. Rousseau and L. Desgranges and F. Charlot and N. Millot and J.C. Nièpce and M. Pijolat and F. Valdivieso and G. Baldinozzi and J.F. Bérar},
}

@article{ZALKIND2017202,
title = {Uranium oxidation kinetics monitored by in-situ X-ray diffraction},
journal = {Journal of Nuclear Materials},
volume = {485},
pages = {202-206},
year = {2017},
issn = {0022-3115},
doi = {https://doi.org/10.1016/j.jnucmat.2016.12.021},
url = {https://www.sciencedirect.com/science/article/pii/S002231151630318X},
author = {S. Zalkind and G. Rafailov and I. Halevy and T. Livneh and A. Rubin and H. Maimon and D. Schweke},
}

@article{MISTARIHI2023154417,
title = {The oxidation of uranium diboride in flowing air atmospheres},
journal = {Journal of Nuclear Materials},
volume = {580},
pages = {154417},
year = {2023},
issn = {0022-3115},
doi = {https://doi.org/10.1016/j.jnucmat.2023.154417},
url = {https://www.sciencedirect.com/science/article/pii/S002231152300185X},
author = {Q. Mistarihi and F. Martini and J. Buckley and S.C. Middleburgh and T.J. Abram and J. Turner},
}

@article{Jolkkonen04052017,
author = {Mikael Jolkkonen and Pertti Malkki and Kyle Johnson and Janne Wallenius},
title = {Uranium nitride fuels in superheated steam},
journal = {Journal of Nuclear Science and Technology},
volume = {54},
number = {5},
pages = {513--519},
year = {2017},
publisher = {Taylor \& Francis},
doi = {10.1080/00223131.2017.1291372},
URL = {  
        https://doi.org/10.1080/00223131.2017.1291372
},
eprint = { 
    
        https://doi.org/10.1080/00223131.2017.1291372
}
}

@article{GASPARRINI2024154944,
title = {On the oxidation and ignition of uranium carbide fragments in air and comparison with zirconium carbide oxidation},
journal = {Journal of Nuclear Materials},
volume = {592},
pages = {154944},
year = {2024},
issn = {0022-3115},
doi = {https://doi.org/10.1016/j.jnucmat.2024.154944},
url = {https://www.sciencedirect.com/science/article/pii/S0022311524000473},
author = {C. Gasparrini and R. Podor and O. Fiquet and M.J.D. Rushton and W.E. Lee}
}

@article{TURNER2022153471,
title = {UN-UB2 Composite fuel material; improved water tolerance with integral burnable absorber},
journal = {Journal of Nuclear Materials},
volume = {559},
pages = {153471},
year = {2022},
issn = {0022-3115},
doi = {https://doi.org/10.1016/j.jnucmat.2021.153471},
url = {https://www.sciencedirect.com/science/article/pii/S0022311521006917},
author = {J. Turner and J. Buckley and R.N. Worth and M. Salata-Barnett and M.J.J. Schmidt and T.J. Abram},
}

@article{GONCHAROV2023154446,
title = {Energetics of oxidation and formation of uranium monocarbide},
journal = {Journal of Nuclear Materials},
volume = {581},
pages = {154446},
year = {2023},
issn = {0022-3115},
doi = {https://doi.org/10.1016/j.jnucmat.2023.154446},
url = {https://www.sciencedirect.com/science/article/pii/S0022311523002143},
author = {Vitaliy G. Goncharov and Juejing Liu and Andrew Strzelecki and Arjen {van Veelen} and Chris Benmore and Hakim Boukhalfa and Joshua T. White and Hongwu Xu and Xiaofeng Guo},
}

@techreport{Bhowmik2023MultiLevelIrradiation,
  author       = {Bhowmik, Palash Kumar and Sabharwall, Piyush and Heidrich, Brenden J. and Howard, Richard},
  title        = {Accelerating Nuclear Fuels and Materials Qualification by Multi-Level Irradiation Experiment Campaign},
  institution  = {Idaho National Laboratory (INL)},
  address      = {Idaho Falls, ID, United States},
  year         = {2023}
}

@article{THOMAS2020152161,
title = {The application of synchrotron micro-computed tomography to characterize the three-dimensional microstructure in irradiated nuclear fuel},
journal = {Journal of Nuclear Materials},
volume = {537},
pages = {152161},
year = {2020},
issn = {0022-3115},
doi = {https://doi.org/10.1016/j.jnucmat.2020.152161},
url = {https://www.sciencedirect.com/science/article/pii/S0022311519310104},
author = {Jonova Thomas and Alejandro {Figueroa Bengoa} and Sri Tapaswi Nori and Ran Ren and Peter Kenesei and Jon Almer and James Hunter and Jason Harp and Maria A. Okuniewski},
}

@article{DEGUELDRE2016242,
title = {Post irradiation examination of nuclear fuel: Toward a complete analysis},
journal = {Progress in Nuclear Energy},
volume = {92},
pages = {242-253},
year = {2016},
issn = {0149-1970},
doi = {https://doi.org/10.1016/j.pnucene.2016.03.025},
url = {https://www.sciencedirect.com/science/article/pii/S0149197016300713},
author = {Claude Degueldre and Johannes Bertsch and Matthias Martin}
}

@Article{Lang2015,
author={Lang, Maik
and Tracy, Cameron L.
and Palomares, Raul I.
and Zhang, Fuxiang
and Severin, Daniel
and Bender, Markus
and Trautmann, Christina
and Park, Changyong
and Prakapenka, Vitali B.
and Skuratov, Vladimir A.
and Ewing, Rodney C.},
title={Characterization of ion-induced radiation effects in nuclear materials using synchrotron x-ray techniques},
journal={Journal of Materials Research},
year={2015},
month={05},
day={01},
volume={30},
number={9},
pages={1366-1379},
issn={2044-5326},
doi={10.1557/jmr.2015.6},
url={https://doi.org/10.1557/jmr.2015.6}
}

@techreport{Koenig1960CeramicCermet,
  author       = {Koenig, N. R. and Webb, B. A.},
  title        = {Properties of Ceramic and Cermet Fuels for Sodium Graphite Reactors},
  institution  = {Atomics International},
  year         = {1960}
}

@article{TURNER2020151919,
title = {Steam performance of UB2/U3Si2 composite fuel pellets, compared to U3Si2 reference behaviour},
journal = {Journal of Nuclear Materials},
volume = {529},
pages = {151919},
year = {2020},
issn = {0022-3115},
doi = {https://doi.org/10.1016/j.jnucmat.2019.151919},
url = {https://www.sciencedirect.com/science/article/pii/S0022311519310931},
author = {Joel Turner and Tim Abram},
}

@article{WATKINS2022153502,
title = {Challenges and opportunities to alloyed and composite fuel architectures to mitigate high uranium density fuel oxidation: Uranium diboride and uranium carbide},
journal = {Journal of Nuclear Materials},
volume = {560},
pages = {153502},
year = {2022},
issn = {0022-3115},
doi = {https://doi.org/10.1016/j.jnucmat.2021.153502},
url = {https://www.sciencedirect.com/science/article/pii/S0022311521007224},
author = {Jennifer K. Watkins and Adrian R. Wagner and Adrian Gonzales and Brian J. Jaques and Elizabeth S. Sooby}
}

@article{Ishizawa:a18575,
author = "Ishizawa, N. and Miyata, T. and Minato, I. and Marumo, F. and Iwai, S.",
title = "{A structural investigation of {$\alpha$}-Al${\sb 2}$O${\sb 3}$ at 2170 K}",
journal = "Acta Crystallographica Section B",
year = "1980",
volume = "36",
number = "2",
pages = "228--230",
month = "2",
doi = {10.1107/S0567740880002981},
url = {https://doi.org/10.1107/S0567740880002981},
}

@article{Beckman1956,
  author  = {Beckman, Gunvor and Kiessling, Roland},
  title   = {Thermal Expansion Coefficients for Uranium Boride and $\beta$-Uranium Silicide},
  journal = {Nature},
  year    = {1956},
  volume  = {178},
  number  = {4546},
  pages   = {1341--1341},
  doi     = {10.1038/1781341a0},
  url     = {https://doi.org/10.1038/1781341a0},
}

\end{document}